\pdfoutput=1

\documentclass[11pt,twoside,a4paper,cmspaper,final,collab]{cms-tdr}
\def\svnVersion{303687}\def\cmsCernNo{2013-037}\def\cmsCernDate{\today}\def\cmsMessage{Published in Physics Letters B as \href{http://dx.doi.org/10.1016/j.physletb.2015.07.065}{\doi{10.1016/j.physletb.2015.07.065.}}}

\begin{document}\cmsNoteHeader{SMP-13-013}

\hyphenation{had-ron-i-za-tion}
\hyphenation{cal-or-i-me-ter}
\hyphenation{de-vices}
\RCS$Revision: 303655 $
\RCS$HeadURL: svn+ssh://svn.cern.ch/reps/tdr2/papers/SMP-13-013/trunk/SMP-13-013.tex $
\RCS$Id: SMP-13-013.tex 303655 2015-09-14 13:45:25Z mstoye $

\providecommand{\qt}{\ensuremath{q_\mathrm{T}}\xspace}

\cmsNoteHeader{SMP-13-013}
\title{Measurement of the Z boson differential cross section in transverse momentum and rapidity in proton-proton collisions at 8\TeV}

\date{\today}

\abstract{
We present a measurement of the Z boson differential cross section
 in rapidity and transverse momentum using
a data sample of pp collision events at a centre-of-mass energy $\sqrt{s}=8$\TeV, corresponding
to an integrated luminosity of 19.7\fbinv. The Z boson is
identified via its decay to a pair of muons. The measurement provides a
precision test of quantum chromodynamics over a large region of phase space. In
addition, due to the small experimental uncertainties in the measurement the data has the potential to constrain the gluon parton distribution function in the kinematic regime important for Higgs boson production via gluon
fusion. The results agree with the next-to-next-to-leading-order predictions computed with the \FEWZ program.
The results are also compared to the commonly used leading-order \MADGRAPH and next-to-leading-order \POWHEG generators.
}

\hypersetup{%
pdfauthor={CMS Collaboration},%
pdftitle={Measurement of the Z boson differential cross section in transverse momentum and rapidity in proton-proton collisions at 8 TeV},%
pdfsubject={CMS},%
pdfkeywords={CMS, physics, Z boson differential cross section}}

\maketitle
\section{Introduction}
The production of lepton pairs in proton-proton collisions is dominated
by the Drell--Yan (DY) process i.e. the production of an intermediate $\gamma^*/$Z boson by the incoming partons. Measurements of the cross sections
as a function of the mass of the intermediate boson (hereafter referred to as the
`Z boson'), rapidity, and transverse momentum provide a
very sensitive test of quantum chromodynamics (QCD). Precise measurement of
the differential cross section also allows comparisons to calculations employing different parton
distribution functions (PDF) and underlying theoretical models. Finally, the understanding of DY lepton pair production is
important in the study of several physics processes, such as diboson
and \ttbar production, as well as in searches for new resonances
decaying to dileptons in models of physics beyond the standard model.
Differential measurements of Z boson production at the LHC have already
been performed~\cite{Chatrchyan:2011wt,Aad:2011gj,Aad:2014xaa,Aaij:2012mda,Aaij:2012vn,Chatrchyan:2013tia,CMS:2014jea,Aad:2012wfa}.

In this Letter we present the first measurement of the DY cross
section at a centre-of-mass energy of 8\TeV for dimuon pairs in the vicinity of the Z boson peak, doubly
differential in the transverse momentum $\qt$ and in the rapidity
$y$ of the Z boson. The analysis uses the data sample of pp collisions collected with the CMS detector at the LHC in 2012, corresponding to an integrated luminosity of 19.7\fbinv. We present the absolute fiducial cross section and the fiducial cross section normalised to the inclusive fiducial cross section.  The measurement probes the production of Z
bosons up to high transverse momenta, $\qt\sim 100$\GeV, a
kinematic regime where the production is dominated by gluon-quark
fusion. The precision of this measurement leads to experimental uncertainties smaller than or similar to the uncertainties of the gluon PDF in the kinematic region that is relevant to
the production of the Higgs boson via the gluon fusion mechanism.
Using the Z boson production process to constrain the gluon PDF~\cite{PDFgluon} in the future would be complementary to other processes such as direct
photon production~\cite{d'Enterria:2012yj} and top-quark pair
production~\cite{Czakon:2013tha} that constrain the gluon PDF in this
regime.
Moreover, several of the experimental systematic uncertainties
in the DY measurement are uncorrelated with these other processes.  The latter  have more complex topologies and thus have complementary and potentially larger
systematic uncertainties.

\section{The CMS detector}
 A more detailed description of the CMS detector, together with a definition of the coordinate system and the relevant kinematic variables, can be found in Ref.~\cite{Chatrchyan:2008aa}. The central feature of the CMS apparatus is a superconducting solenoid
of 6\unit{m} internal diameter, providing a magnetic field of
3.8\unit{T}. Within the solenoid volume are a silicon
pixel and strip tracker, a lead tungsten crystal electromagnetic
calorimeter (ECAL), and a brass/scintillator hadron calorimeter (HCAL),
each composed of a barrel and two endcap sections. Muons are measured
in gas-ionisation detectors embedded in the steel flux-return yoke
outside the solenoid. Extensive forward calorimetry complements the
coverage provided by the barrel and endcap detectors.  Muons are
measured in the pseudorapidity range $\abs{\eta}< 2.4$, with detection
planes made using three technologies: drift tubes, cathode strip
chambers, and resistive-plate chambers. Matching muons to tracks
measured in the silicon tracker results in a relative \pt resolution
 of 1.3--2.0\% in the barrel and better than 6\% in the endcaps, for muons with $20 <\pt < 100$\GeV. The \pt resolution in the barrel is
better than 10\% for muons with \pt up to
1\TeV~\cite{Chatrchyan:2012xi}.  The particle-flow event
reconstruction~\cite{CMS:2009nxa,CMS:2010byl} is used in
this analysis. It works by reconstructing and identifying each
particle with an optimised combination of all subdetector
information. The energy of photons is directly obtained from the ECAL
measurement, corrected for zero-suppression effects. The energy of
electrons is determined from a combination of the track momentum at the
main interaction vertex, the corresponding ECAL cluster energy, and the
energy sum of all bremsstrahlung photons. The
energy of muons is obtained from the corresponding track momentum. The
energy of charged hadrons is determined from a combination of the track
momentum and the corresponding ECAL and HCAL energies, corrected for
zero-suppression effects, and calibrated for the nonlinear response of
the calorimeters. Finally, the energy of neutral hadrons is obtained from
the corresponding calibrated ECAL and HCAL energies.  The first level
of the CMS trigger system, composed of custom hardware processors, uses
information from the calorimeters and muon detectors to select the most
interesting events in a fixed time interval of less than 4\mus. The high-level trigger processor farm further decreases the event rate from
around 100\unit{kHz} to around 400\unit{Hz} before data storage.

\section{Simulation}
The signal process is simulated using the leading-order (LO) \MADGRAPH~1.3.30~\cite{Alwall:2011uj} generator with 0--4 additional jets,
interfaced with \PYTHIA~\cite{Sjostrand:2006za} v6.4.24 with the Z2* tune~\cite{Chatrchyan:2013gfi}. The matching between matrix element calculation and parton shower is performed with the \kt-MLM algorithm \cite{Alwall:2007fs}. Multiple-parton interactions are accounted for via \PYTHIA. The LO CTEQ6L1~\cite{Pumplin:2002vw} PDF set is used for the generation. As a cross-check, a second signal sample is simulated using the next-to-leading-order (NLO) \POWHEG~\cite{Nason:2004rx,Frixione:2007vw,Alioli:2010xd,Alioli:2008gx} generator interfaced with \PYTHIA. For this generation the NLO CT10~\cite{Lai:2010vv} PDF set
is used.

The backgrounds are generated with \MADGRAPH (W+jets,
\ttbar, $\PGt\PGt$), \POWHEG (single top quark
~\cite{Alioli:2009je,Re:2010bp}), and \PYTHIA (dibosons, WW, WZ,
ZZ). The inclusive cross sections of DY, W+jets~\cite{Gavin:2010az}, and
\ttbar~\cite{Czakon:2013goa} processes are normalised to
next-to-next-to-leading-order (NNLO) predictions. In addition, for the single top quark a
higher-order (approximate NNLO~\cite{Kidonakis:2012db}) inclusive cross section is used. The
generated events are passed through a detector simulation based on
\GEANTfour~\cite{Agostinelli:2002hh}. The simulated processes are overlaid
with minimum bias collisions in order to reproduce the distribution of
the number of additional proton-proton interactions per bunch crossing
(pileup) present in data.

In addition, for comparison with the final result the double differential cross section is computed with
\FEWZ~3.1.b2~\cite{Li:2012wna} at NNLO. The electroweak corrections
are computed at NLO and initial-state photon radiation and photon-induced
processes are included in the generation. The computation is done for
each $\qt$ bin separately. The factorisation and renormalisation
scales are chosen as $\sqrt{\smash[b]{M_\Z^2+\qt^2}}$, where $M_\Z$ is the mass of
the Z boson and $\qt$ is the value of the lower edge of the
corresponding bin in $\qt$. For the computation the NNLO NNPDF23 PDF
set with radiative corrections~\cite{Ball:2013hta} is used.

\section{Event selection}
An isolated single-muon trigger is used with a threshold of $\pt
> 24$\GeV and a requirement of
$\abs{\eta} < 2.1$. The standard CMS baseline offline muon selection~\cite{Chatrchyan:2012xi} is
applied. It requires that the muon candidate is
reconstructed both in the muon detectors and in the inner tracker, with
$\chi^2/n_{\text{dof}}< 10$ for the track fit. In addition, requirements
are placed on the minimum number of pixel and tracker layers that are hit
and on detailed matching criteria between the trajectories reconstructed
in the inner tracker and the muon system. The distance between the muon candidate
trajectory and the primary vertex is required to be smaller than 2\unit{mm}
in the transverse plane and smaller than 5\unit{mm} in the longitudinal
direction. The vertex with the highest sum of $\pt^2$ of associated
tracks is selected as the primary vertex.  The leading reconstructed muon in \pt is required to be the one
selected by the trigger. In order to be within the trigger acceptance, the leading muon is selected with $\pt>25$\GeV and $\abs{\eta} < 2.1$. The second muon is required to have $\pt > 10$\GeV and $\abs{\eta} < 2.4$.
 The relative isolation is defined to be the scalar sum of the transverse momenta of charged hadrons, neutral hadrons, and photons in a cone of $\Delta R=\sqrt{\smash[b]{(\Delta\eta)^2+(\Delta \phi)^2}}<0.4$ around the muon direction, divided by $\pt$. After correction for pileup, the value of the relative isolation is required to be less than 0.12\,(0.5) for the leading (second) muon. A pair of oppositely charged muons is
required to have an invariant mass $M(\mu\mu)$ between 81 and
101\GeV. In the rare case of ambiguity among several reconstructed muons, the muon pair with the invariant
mass closest to the Z boson mass is selected. The absolute rapidity
$\abs{y}$ of the muon pair must be less than 2.

Scale factors are applied to account for known differences between data
and simulation. The efficiencies for the tracking,
the trigger, the muon isolation and identification are determined via a ``tag-and-probe''
method~\cite{Khachatryan:2010xn}. The tracking efficiency is measured in
bins of $\eta$. The trigger efficiency is measured in bins of muon
$\pt$ and $\eta$ for positive and negative muons separately. The identification efficiency is
measured in bins of $\pt$ and $\eta$. In particular phase space regions, especially for higher $\qt$, the second muon can often
point opposite to the Z boson in the azimuthal plane. In that
direction, the hadronic activity from the recoil of the Z boson is
enhanced. Thus the second muon is often less isolated than the leading
muon and the isolation depends on the event kinematics. For that reason,
the requirement for the isolation of the second muon is looser than the requirement for the leading muon,
and the efficiency is measured in variables reflecting
the second muon direction with respect to the Z boson. Three variables for the second muon are chosen to measure this
effect on the efficiency in data: the transverse momentum of the
dimuon system $\qt$, the cosine of the polar angle $\cos\theta^*$
and the azimuthal angle $\phi^*$. The two angles are measured in the Z boson rest frame, where
the $z$ axis is the Z boson flight direction. For $\cos(\theta^*) = -1$ the leptons are more likely to be close to the hadronic recoil. The azimuthal angle is chosen to be zero for the proton closest to the $z$ axis in this frame.
 The isolation efficiency for the leading muon is measured in bins of $\pt$ and $\eta$. These efficiencies are measured in data and
simulation, and scale factors are applied to the simulation to
account for differences with respect to the data.

The backgrounds are small relative to the signal (at the percent level or smaller) and can be divided into two categories: those where the leptons come from Z boson decays and those where the leptons stem from other sources.
 The backgrounds from \ttbar, $\PGt\PGt$, WW, tW, and W+jets are estimated
from specific data samples. Backgrounds typically have two prompt leptons, although not necessarily of the same lepton flavor: flavor universality is used for the background
estimation. The estimation consists of two steps. First, the oppositely charged mixed lepton
$\Pe\PGm$ yields are measured in both data and MC.  Then the ratio of the
yields in data and simulation in this data sample ($\Pe\PGm$ channel) is
 used to normalise the simulation in the muon channel. The
$\Pe\PGm$-channel selection uses the same trigger as the final sample, thus
the same trigger efficiency scale factor is used. In addition, the tracking, the identification, and the isolation efficiency scale factors for the leading
muon are applied. Electrons are selected if they have $\pt(\Pe) > 20$\GeV and $\abs{\eta(\Pe)} < 2.1$, which is similar to the fiducial regions of the muon selection.
 No data-to-simulation correction factors are applied to the electron
identification since the effect on the final results is negligible.
 In order to enhance the statistical precision, the invariant mass range of the $\Pe\PGm$ pairs is increased to [60,120]\GeV. Within the uncertainties, no significant trend in $\qt$ and $\abs{y}$ is observed in the ratio of the
$\Pe\PGm$ yields in data and simulation, and a constant scale factor of $0.987\pm0.008$ is used.
 The WZ and ZZ backgrounds, which
include a true $\Z\to\PGm\PGm$ decay, are taken from
simulation.

\section{Measured observables and granularity}
\label{sec:meas}
The reconstructed and background-corrected double differential distribution in $\qt$
and $\abs{y}$ is unfolded to pre-final-state radiation (FSR) lepton kinematics. The unfolding is
performed to the kinematic region $81\leq M(\mu\mu)<101$\GeV and within the kinematic selection of the leading (second)
muon, $\pt>25(10)$\GeV and $\abs{\eta}<2.1\,(2.4)$. The unfolding
is done using an iterative unfolding technique~\cite{D'Agostini:1994zf}
implemented in the RooUnfold package~\cite{Adye:2011gm}. The bins in
$\qt$ are [0,20], [20,40], [40,60], [60,80], [80,100], [100,120],
[120,140], [140,170], [170,200], [200,1000]. In $\abs{y}$ a constant bin width of 0.4 is used and the binning ends at 2.

\MADGRAPH is used as simulation input to the unfolding. The unfolding is validated by treating the simulated \POWHEG
signal sample as data. The unfolded \POWHEG distribution is found
to be compatible with the  distribution at the generator level within unfolding uncertainty.

\section{Systematic uncertainties}
The sources of systematic uncertainty are ordered by their average size starting with the largest one.
The full covariance matrix is computed for both the normalised and the absolute cross section.
\begin{itemize}
\item Luminosity uncertainty:\\
 The uncertainty in the measurement of the integrated luminosity is 2.6\%~\cite{CMS:2013gfa}.
\item Tracking, muon trigger, isolation, and identification efficiency
correction factors:\\
A potential bias in the measurement of the efficiencies with the tag-and-probe technique is estimated by varying the most sensitive components: the background in simulation is removed and doubled; the signal is parametrised with the sum of two Voigtian functions instead of the sum of a Crystal Ball and a Gaussian function; the efficiencies are parametrised only in $\eta$ but with finer bins; and, only tags with a single available probe are selected for the measurement instead of all possible pairs. For each contribution a 100\% correlation is assumed in the covariance matrix. The effect of statistical uncertainties in the measured data-to-simulation scale factors is estimated by their variation within the uncertainties in a series of pseudo-experiments. Combining the effects extracted from these variations, the systematic uncertainties are typically between 1\% and 1.6\%, depending on the bin, and increase with $\qt$.

\item Pileup uncertainty:\\
The cross section of minimum bias events is varied by ${\pm}5$\% and
the impact of the pileup multiplicity in the simulation on the measurement is used as
correlated uncertainty for all bins. This uncertainty is at maximum around 0.5\% and is negligible compared to the leading uncertainties.
\item Statistical uncertainties of the simulation:\\
The uncertainty due to the limited number of events in
simulation is estimated via pseudo-experiments by varying
the response matrix and the efficiency within the statistical uncertainties.
\item FSR:\\
The simulation is reweighted to reflect the difference between a
soft-collinear approach and the exact O($\alpha$) result, similar to
what was done in Ref.~\cite{Khachatryan:2010xn}. It also reflects effects
from higher-order contributions. The difference between the measurements
with and without the reweighting is assigned as an uncertainty and
is assumed to be fully correlated for the covariance matrix.
\item Backgrounds:
\begin{itemize}
\item \ttbar,tW, WW, W+jets, and $\PGt\PGt$ backgrounds:\\
A 10\% uncertainty is assigned to the scale factor derived in the $\Pe\PGm$ method. This accounts for the statistical uncertainty of the scale factor and for the uncertainties in the lepton efficiencies. For the covariance matrix full correlation is assumed.
\item WZ and ZZ backgrounds:\\
The diboson backgrounds that include a Z boson are determined from simulation. The cross sections are varied by
50\% to estimate the systematic uncertainty. While the inclusive cross sections have been measured to agree reasonably well~\cite{Aad:2012twa,Khachatryan:2014dia,Aad:2012awa}, we assign conservatively 50\% to account for the fact that we use the $\qt$ and $\abs{y}$ shapes from LO calculations.
\end{itemize}
\item Muon momentum resolution:\\
The muon momentum resolution is measured in data and simulation, and
corresponding corrections are applied. The covariance accounting for the
statistical uncertainty of the muon momentum correction measurements is
calculated via pseudo-experiments. In addition, an uncertainty is assigned to take into account possible correlated offsets.
\item Z boson polarisation:\\
The lepton angular distribution of the Drell--Yan process can be described at LO through the coefficients,
 $A_0$--$A_4$~\cite{Collins:1977iv}. However, inaccuracies in the way this is modelled in the simulation can affect the result of the unfolding. The angular coefficients $A_0$--$A_4$ are inferred
in bins of $\qt$ and $\abs{y}$ in~\cite{Zpolarization} for both data and simulation we use. For each
parameter $A_i$ the simulation is independently reweighted to correspond to the data as measured in ~\cite{Zpolarization}. In case the difference in $A_i$ is smaller than the typical theoretical uncertainty of 10\%~\cite{Bern:2011ie} $A_i$ is varied
by 10\%.  The full difference between the default polarisation and the
changed polarisation is assigned as systematic uncertainty. Full
correlation is assumed between the bins.
\item $\qt$ and $y$ shapes:\\
The dependence of the results on the $\qt$ and $y$ shapes of
the simulation is studied by repeating the analysis using \POWHEG
as the signal sample. The results obtained using \MADGRAPH or \POWHEG for the unfolding are compatible with each other within the
statistical uncertainties. In addition, the \MADGRAPH simulation is weighted in fine bins in
$\qt$ and $y$ to match the background-corrected data.
The effect on the result using the reweighted simulation for the unfolding is much smaller than the uncertainties assigned to the limited statistics of simulation and is neglected.
\end{itemize}

\begin{figure*}[!ht]
\centering
\includegraphics[width=0.32\textwidth]{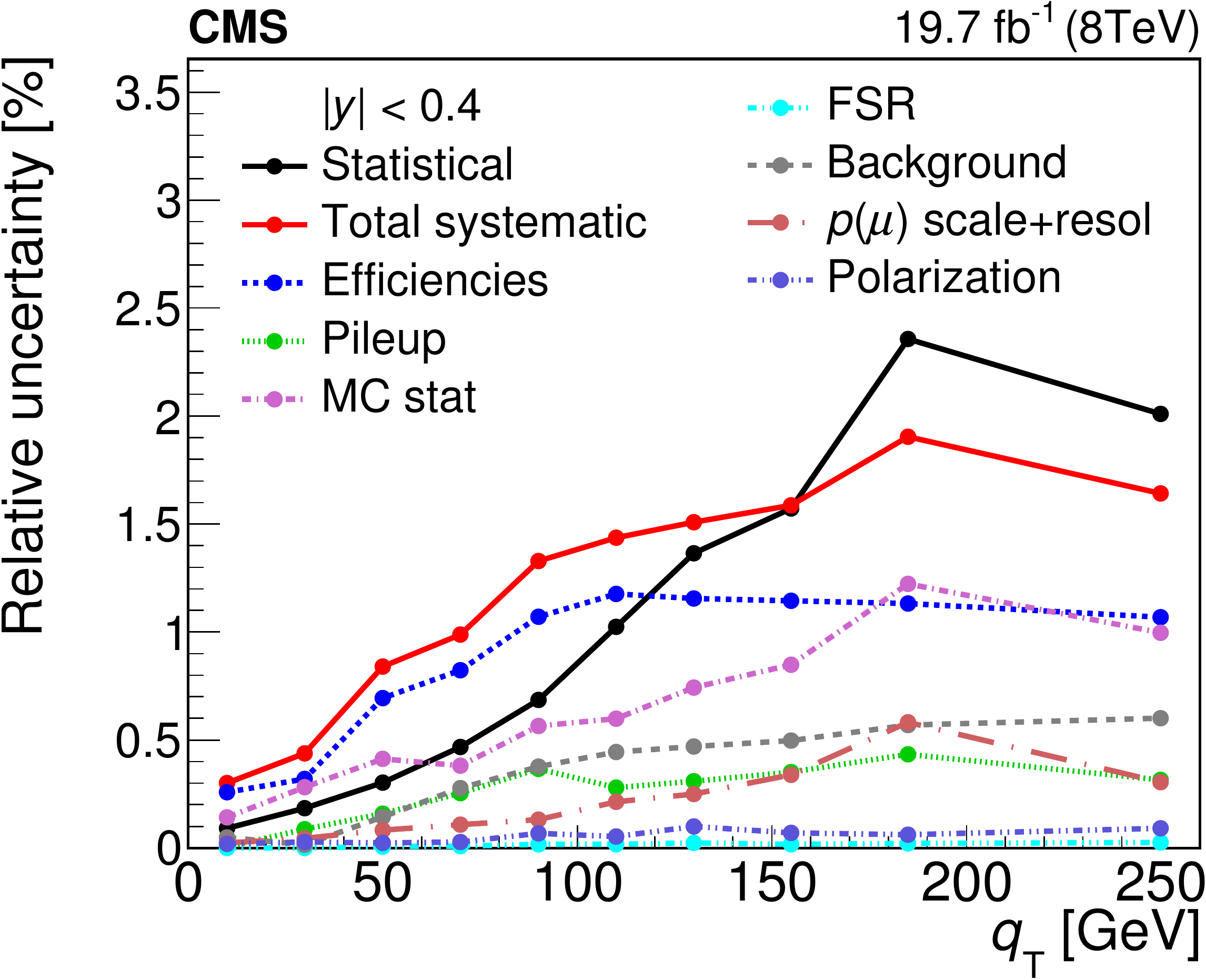}
\includegraphics[width=0.32\textwidth]{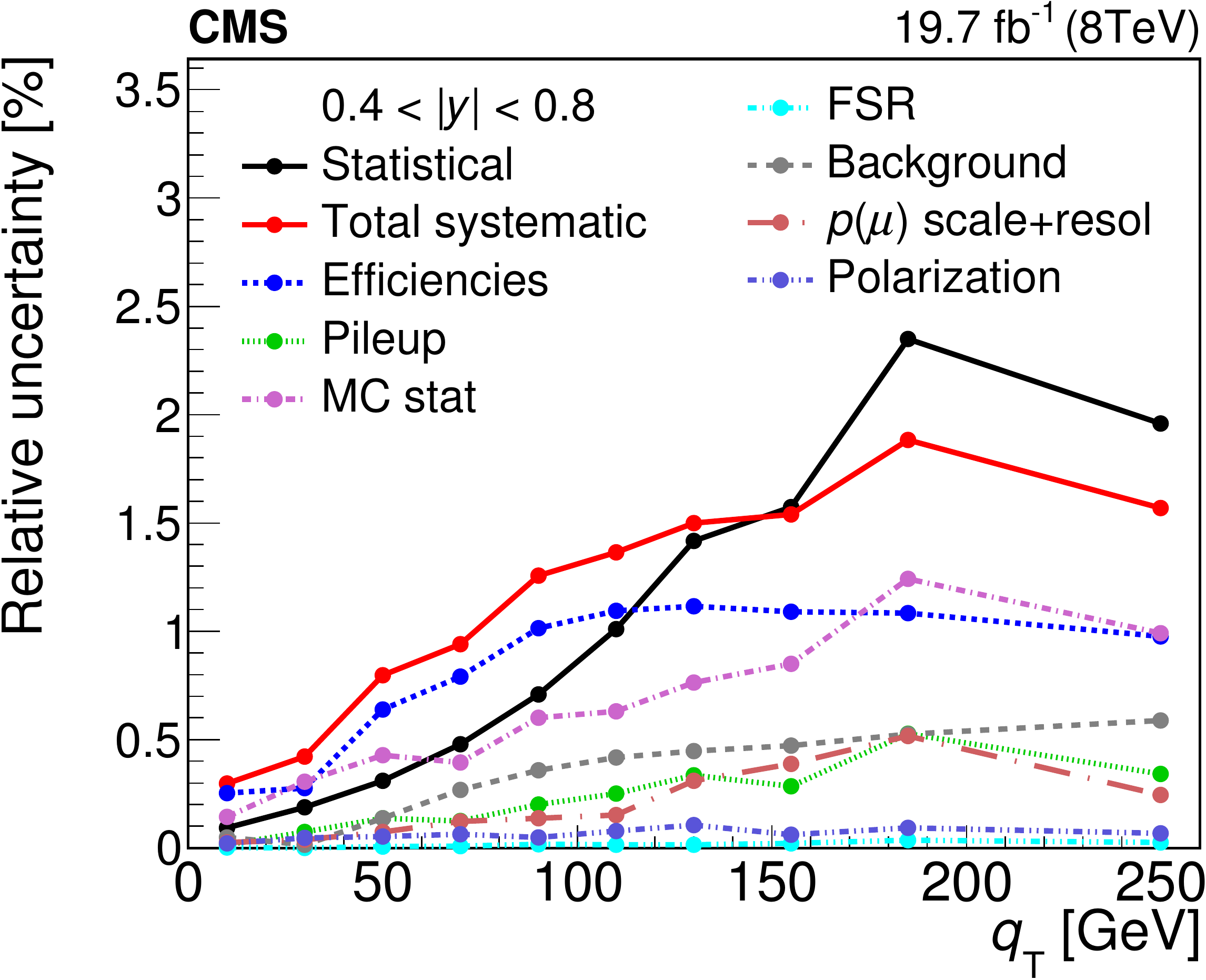}
\includegraphics[width=0.32\textwidth]{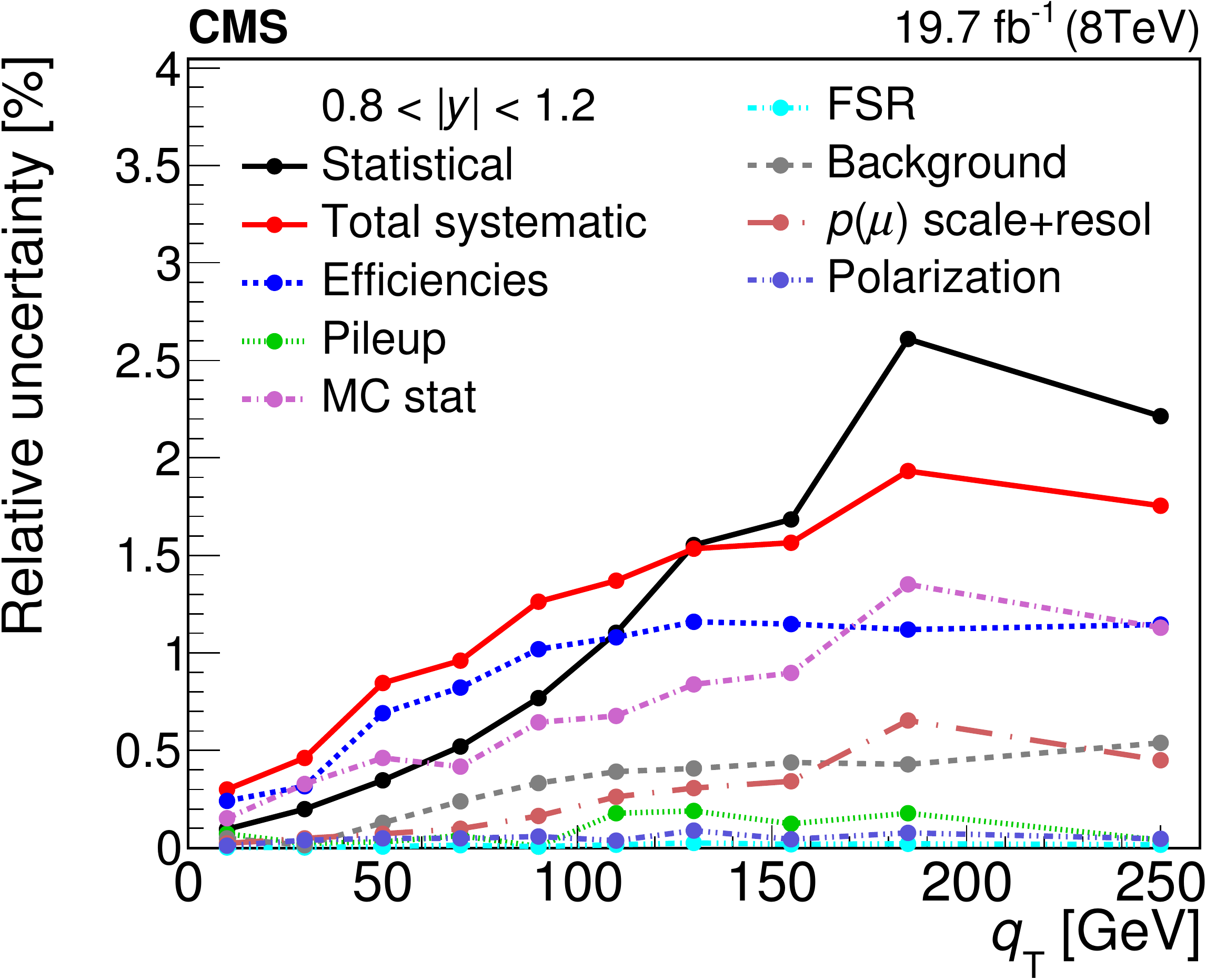}\\
\includegraphics[width=0.32\textwidth]{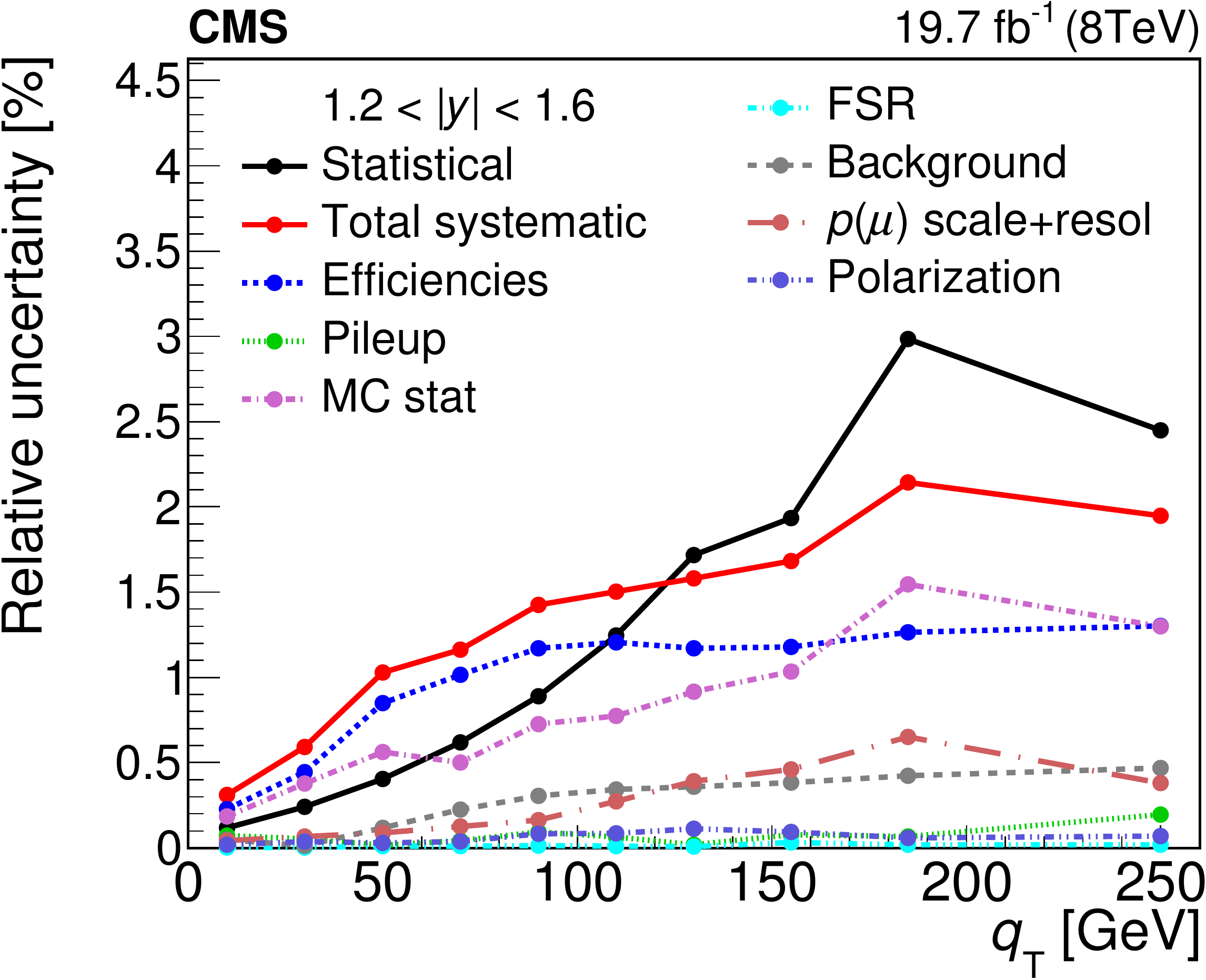}
\includegraphics[width=0.32\textwidth]{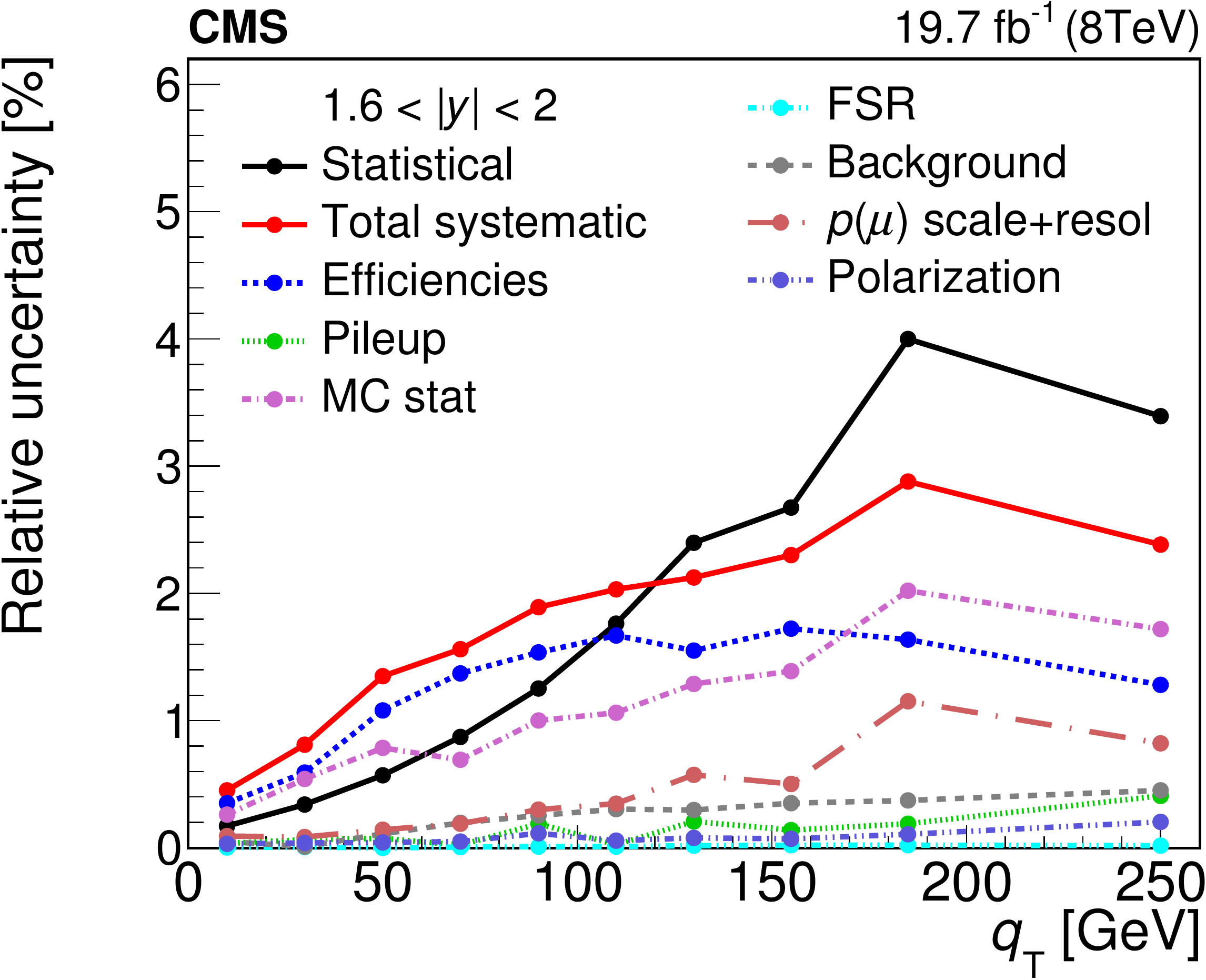}
\includegraphics[width=0.32\textwidth]{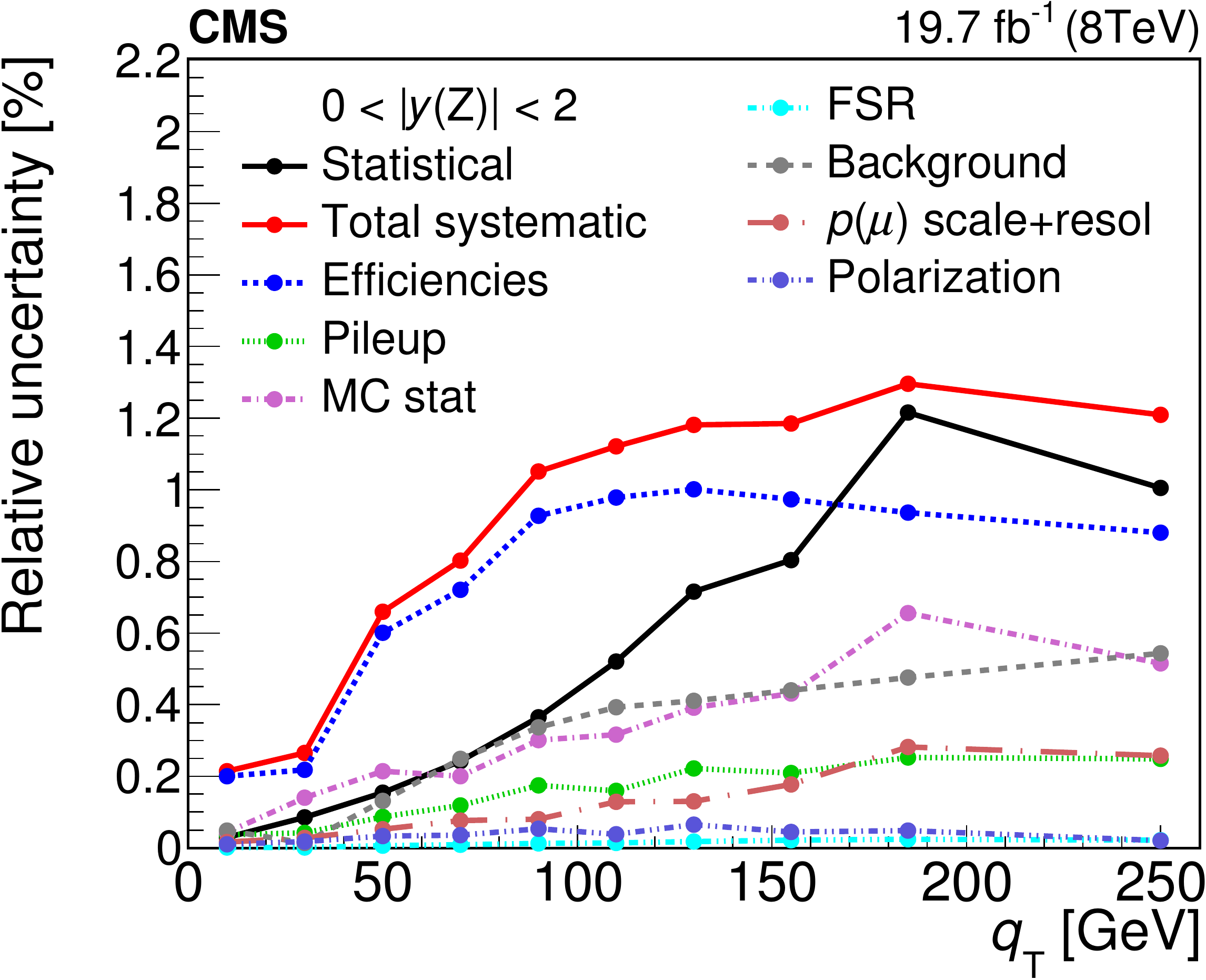}
\caption{Relative uncertainties in percent of the normalised fiducial cross section measurement. Each plot shows the $\qt$ dependence in the indicated ranges of $\abs{y}$.}
\label{fig:unc_normed}
\end{figure*}

\begin{figure*}[!ht]
\centering
\includegraphics[width=0.32\textwidth]{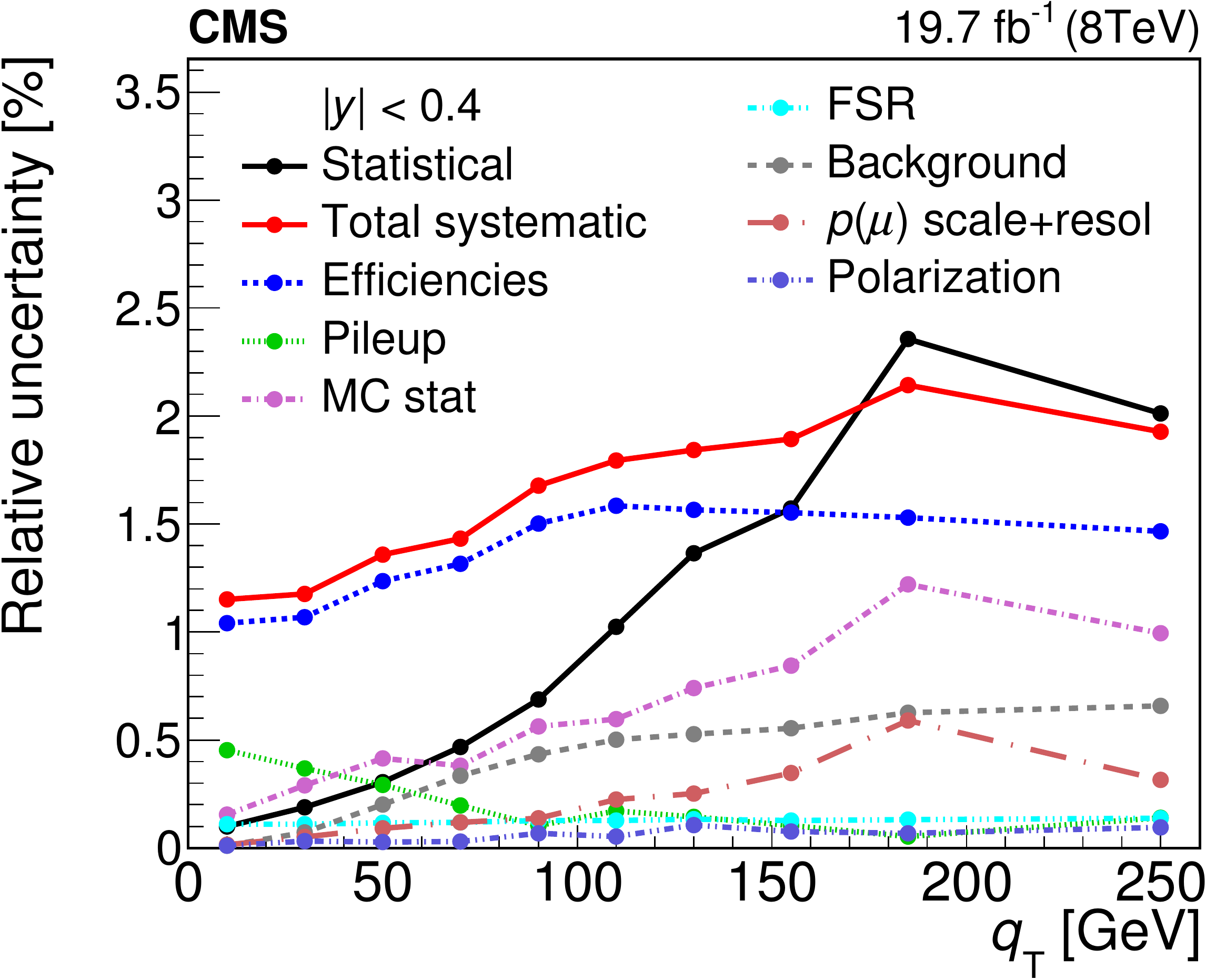}
\includegraphics[width=0.32\textwidth]{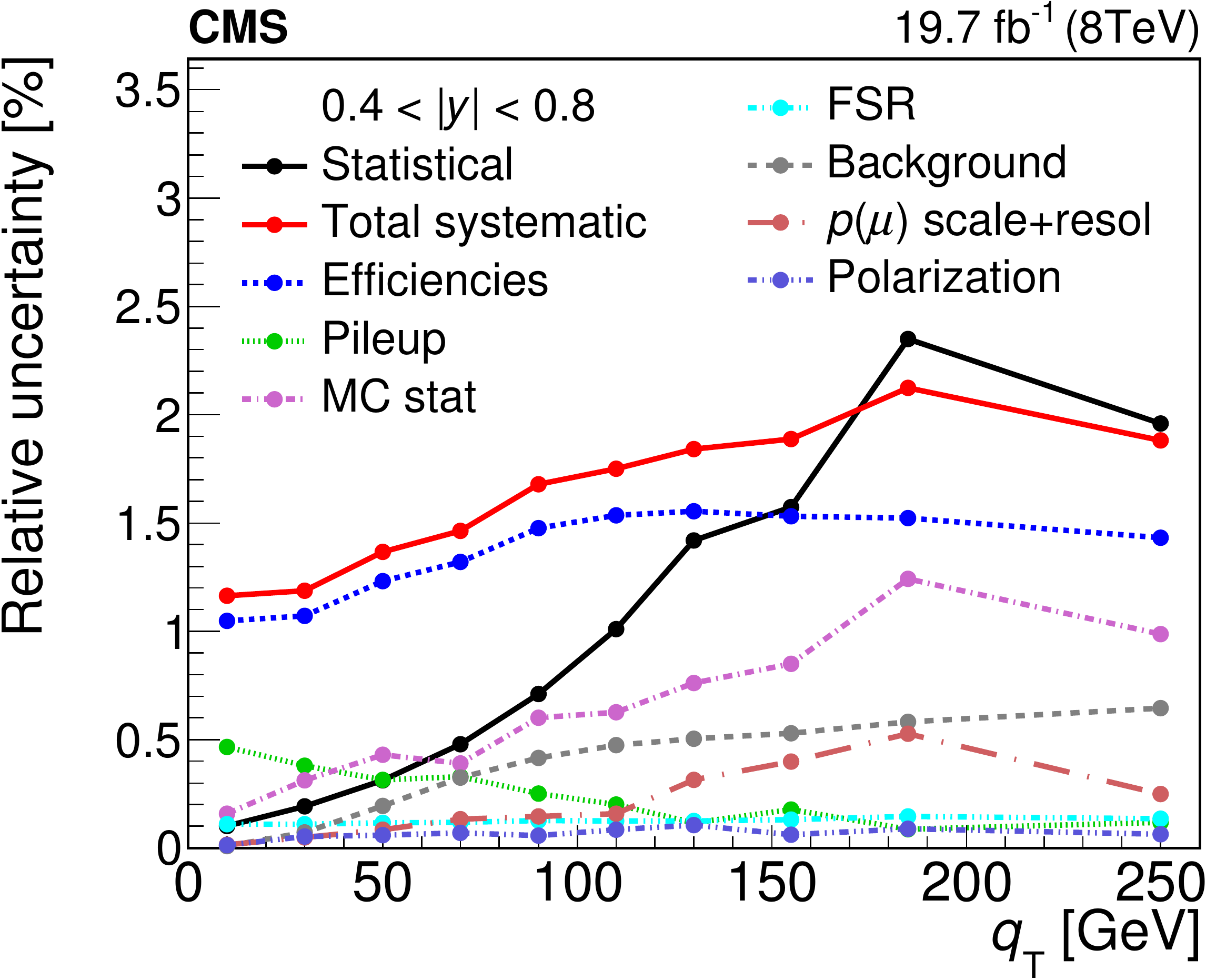}
\includegraphics[width=0.32\textwidth]{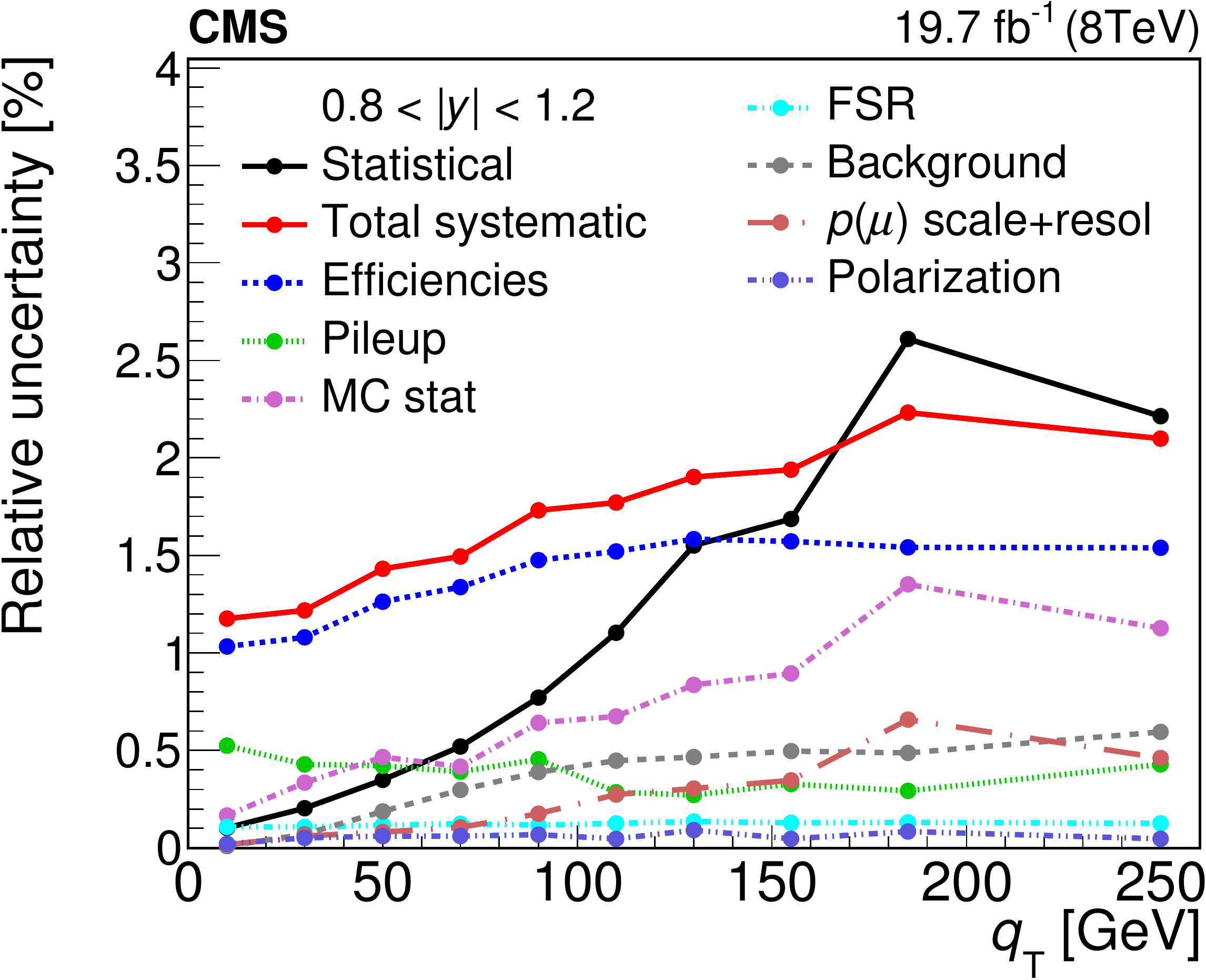}\\
\includegraphics[width=0.32\textwidth]{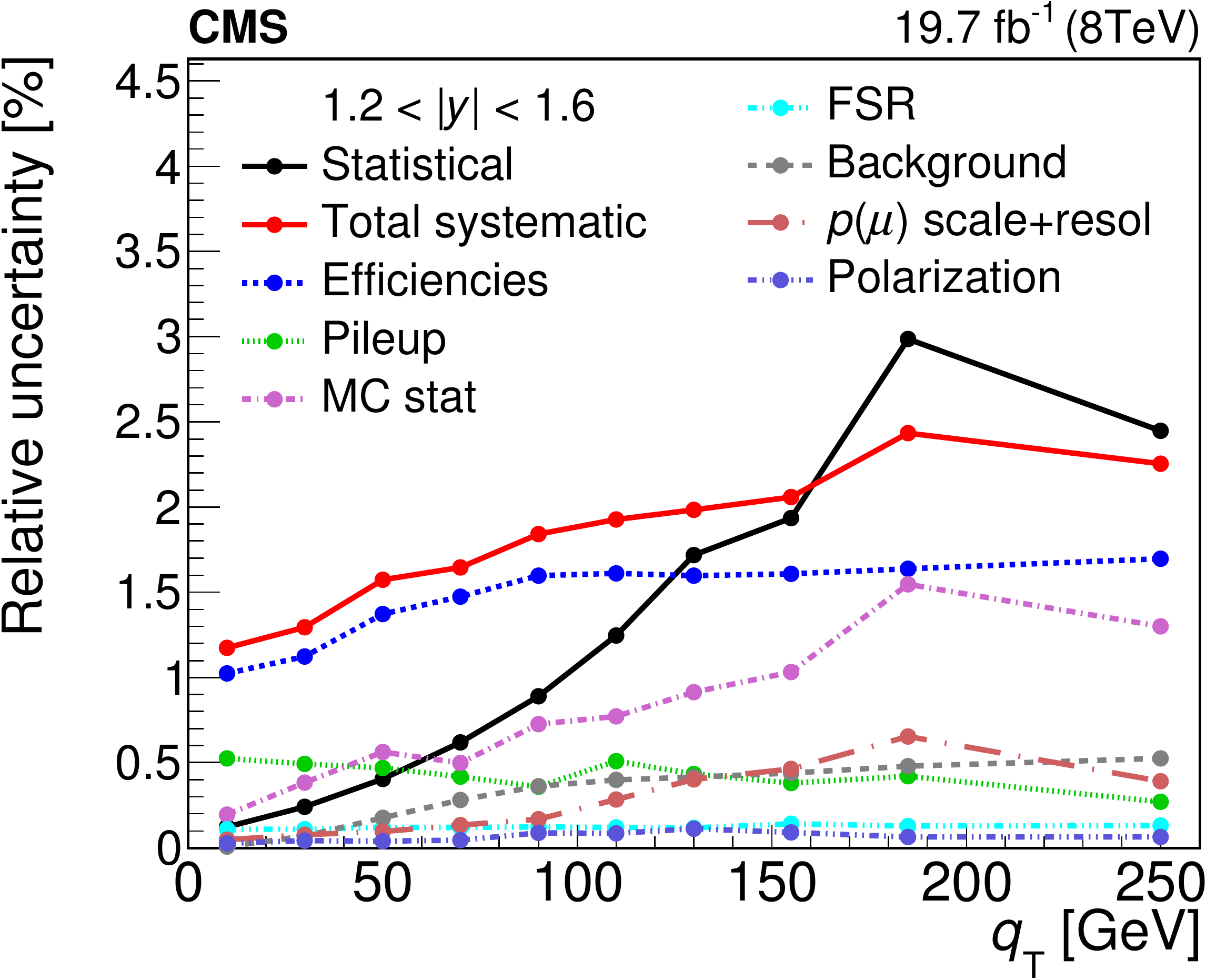}
\includegraphics[width=0.32\textwidth]{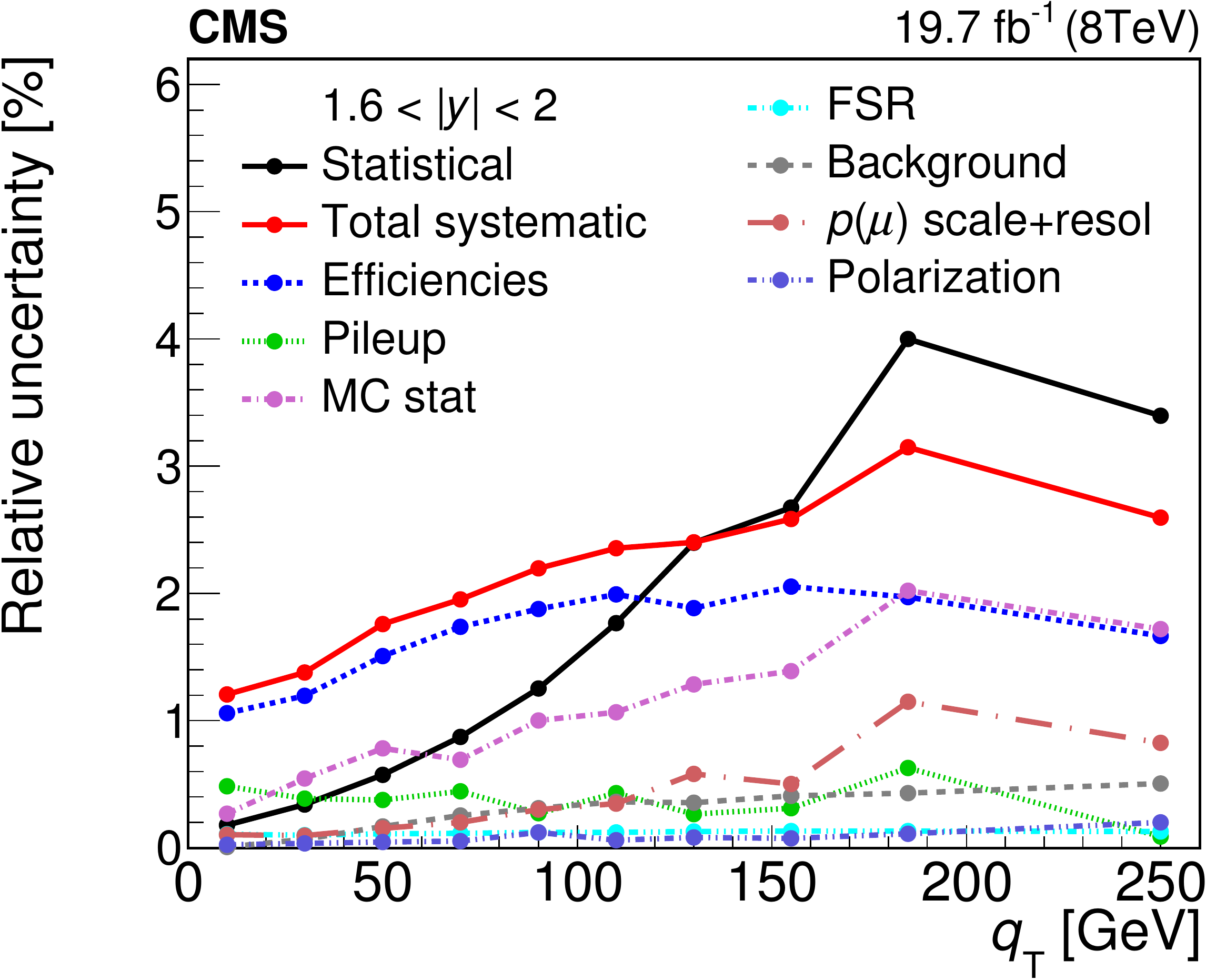}
\includegraphics[width=0.32\textwidth]{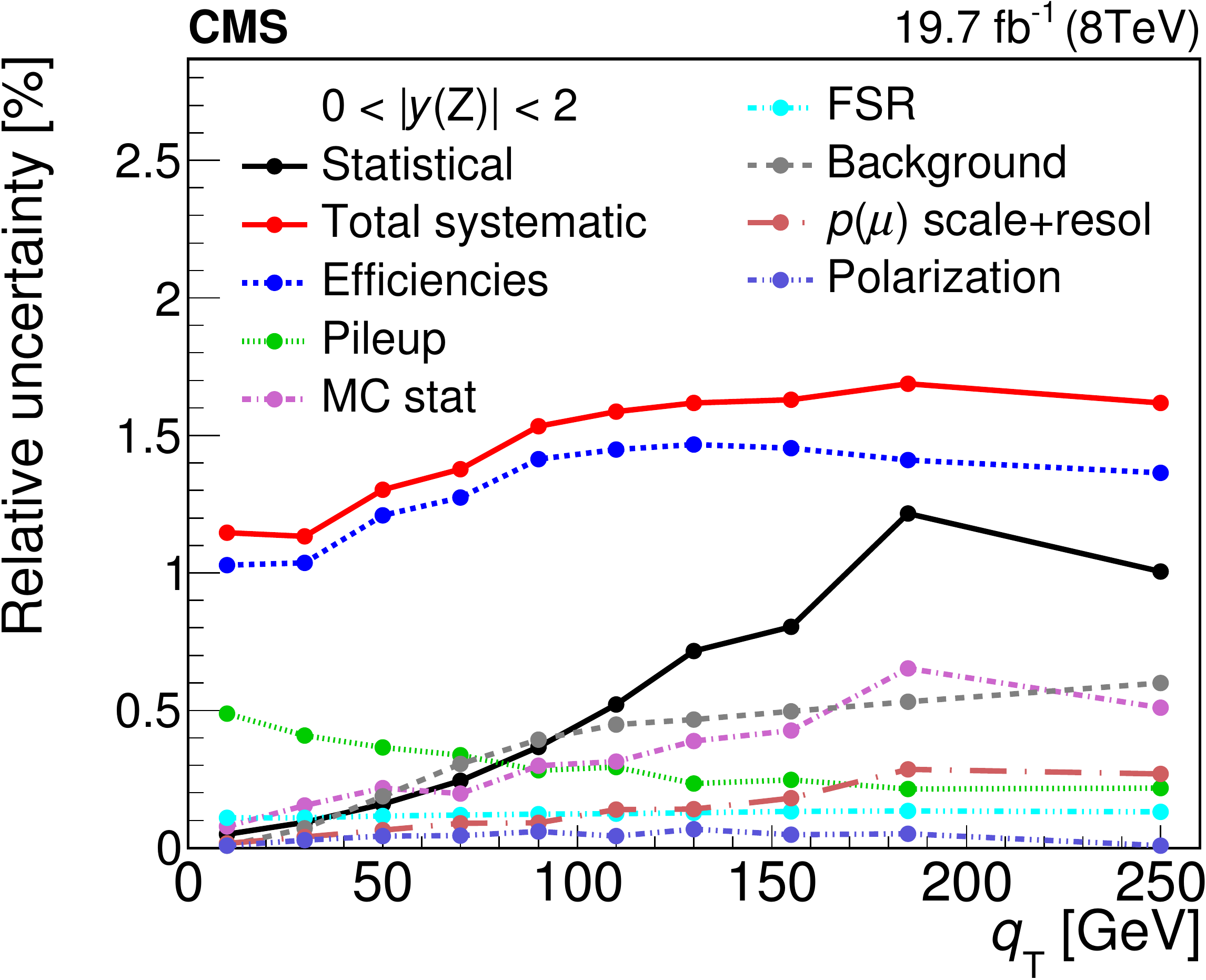}
\caption{Relative uncertainties in percent of the absolute fiducial cross section measurement. The 2.6\% uncertainty in the luminosity is not included. Each plot shows the $\qt$ dependence in the indicated ranges of $\abs{y}$.}
\label{fig:unc_abs}
\end{figure*}

The contributions of the uncertainties to the normalised cross section
measurement are presented in Fig.~\ref{fig:unc_normed}. The systematic uncertainty
is dominated by the uncertainty in the efficiency correction.  In the highest bins of $\qt$ the measurement is dominated by the statistical uncertainties. The uncertainty contributions to the absolute cross section measurement are presented in Fig.~\ref{fig:unc_abs}.

\section{Results}
\begin{figure*}[!ht]
\centering
\includegraphics[width=0.32\textwidth]{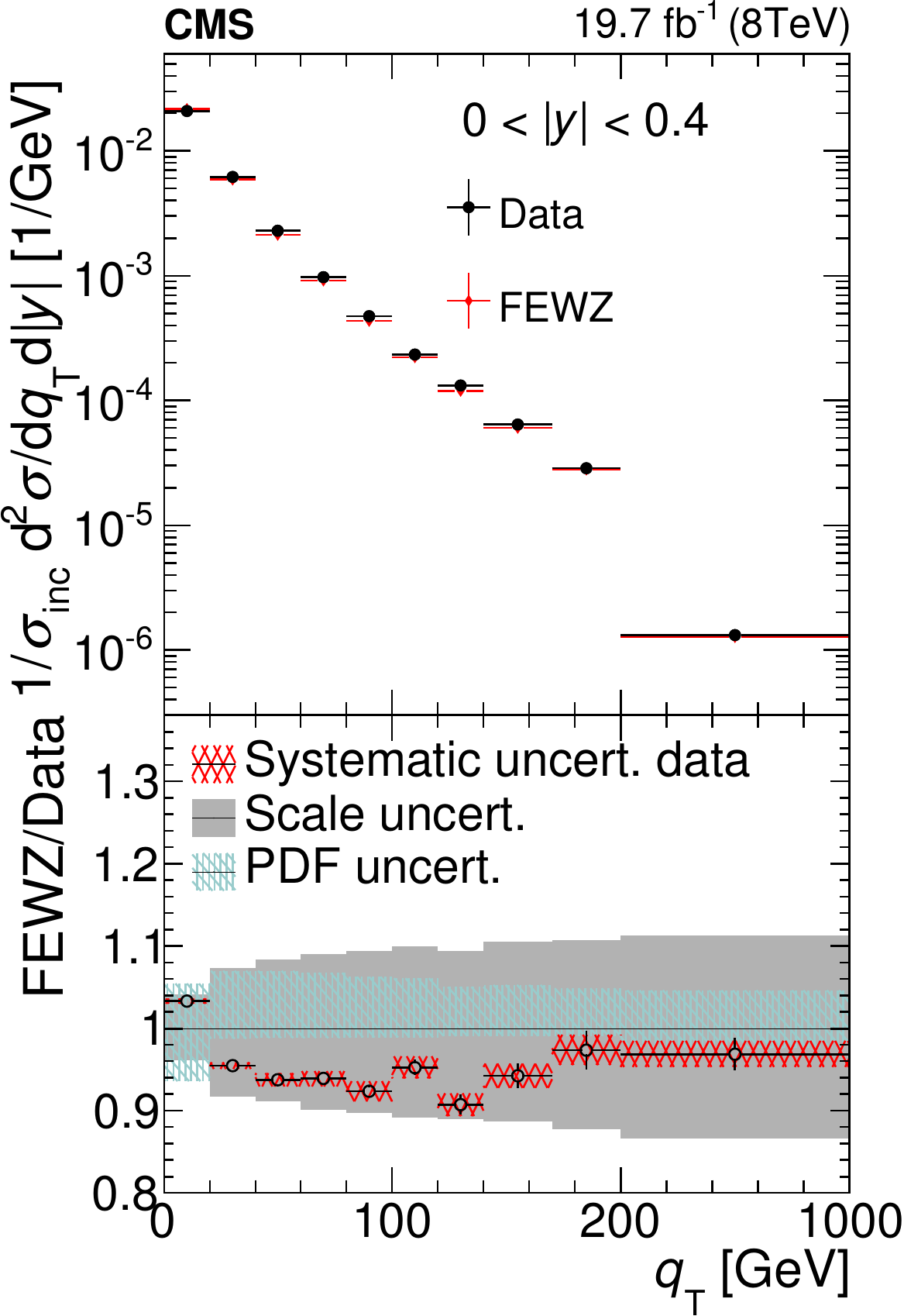}
\includegraphics[width=0.32\textwidth]{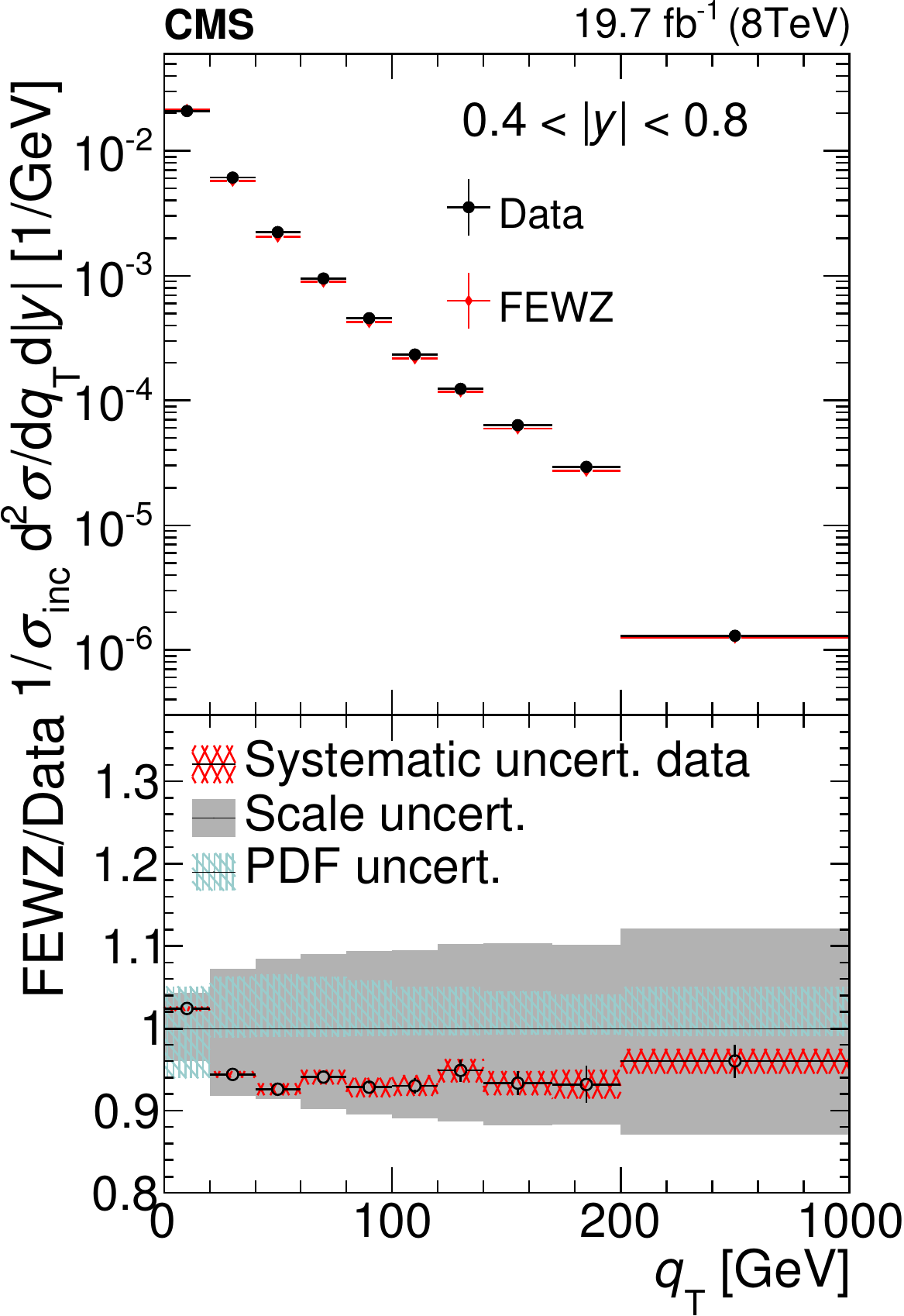}
\includegraphics[width=0.32\textwidth]{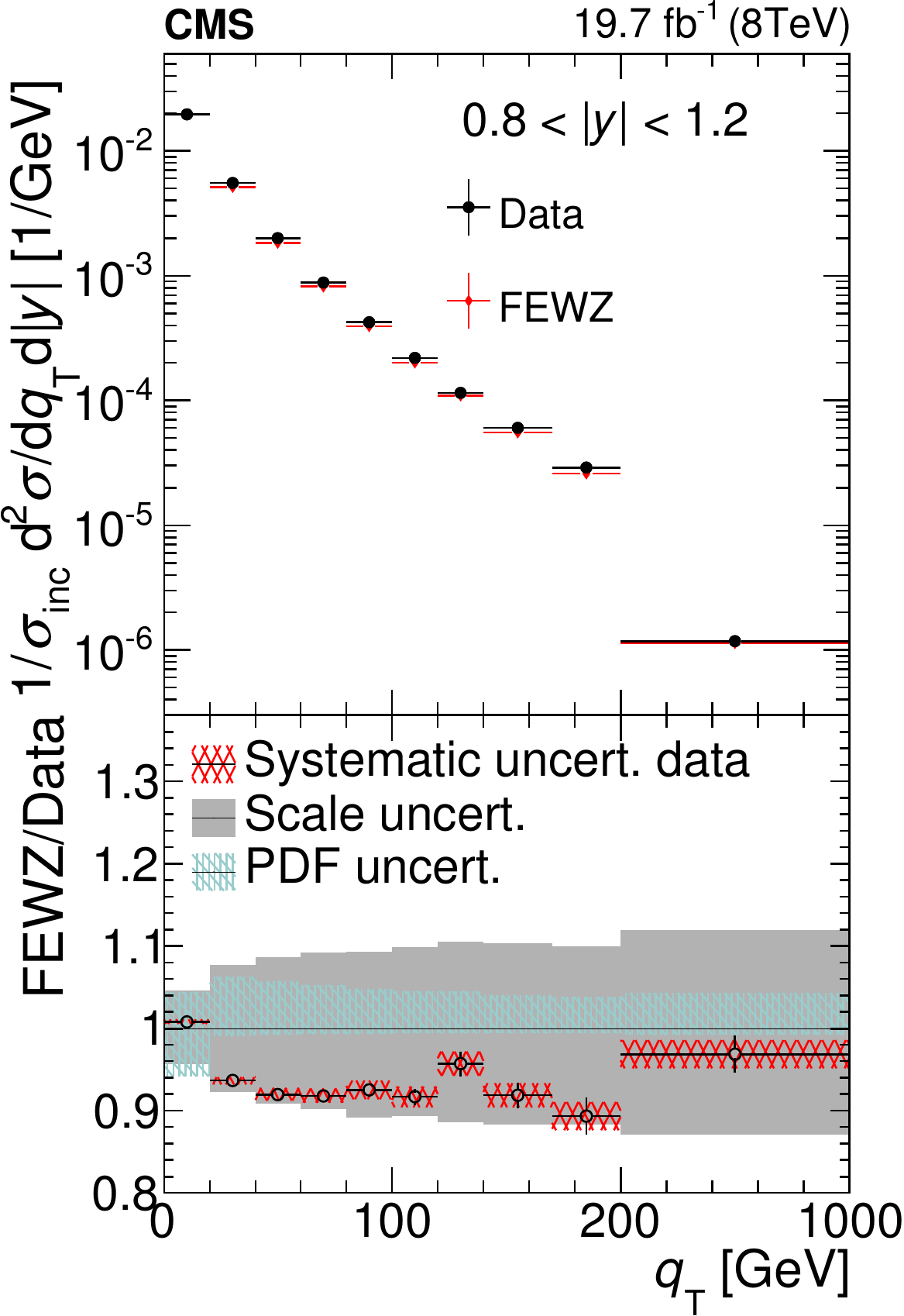}\\
\includegraphics[width=0.32\textwidth]{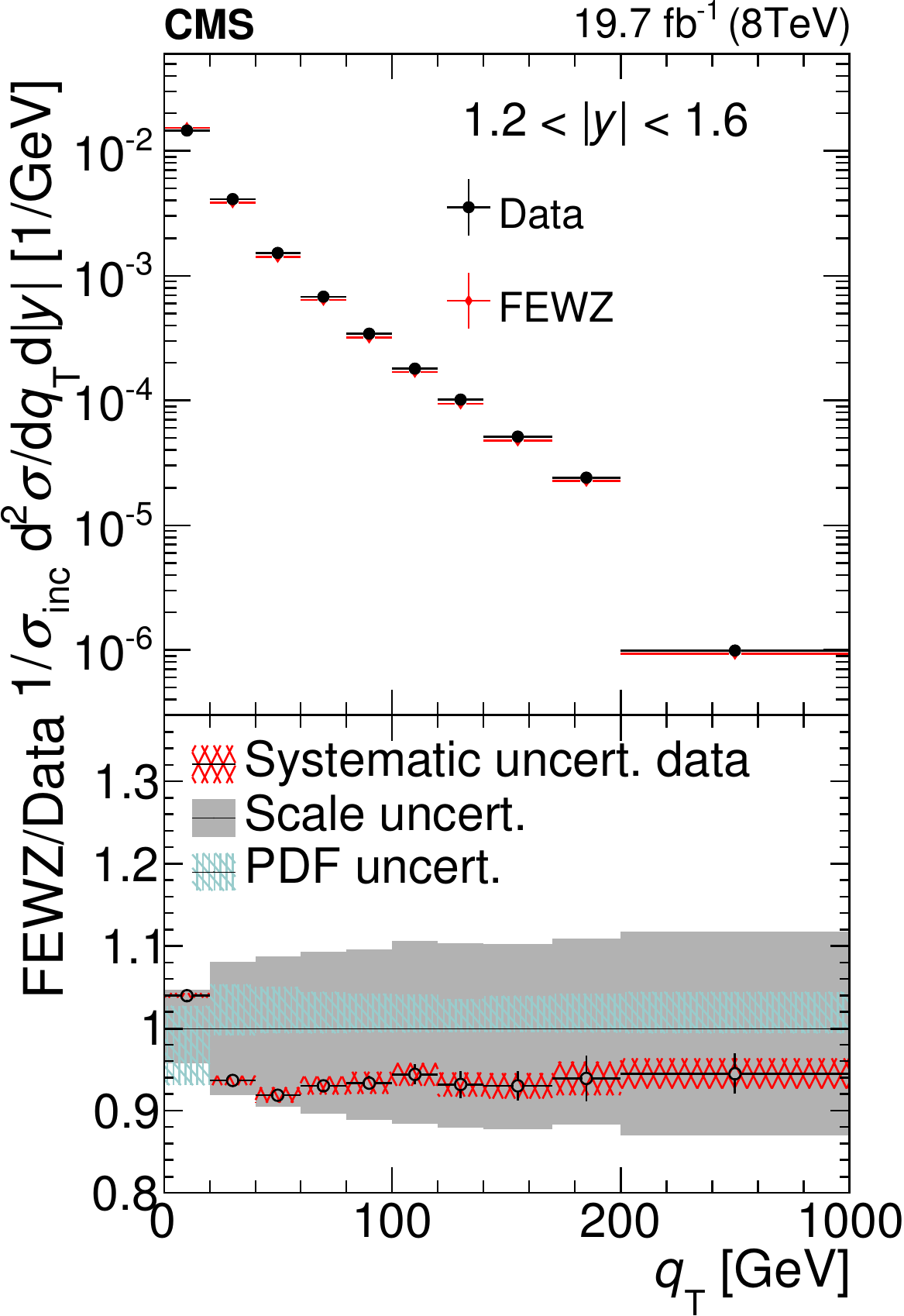}
\includegraphics[width=0.32\textwidth]{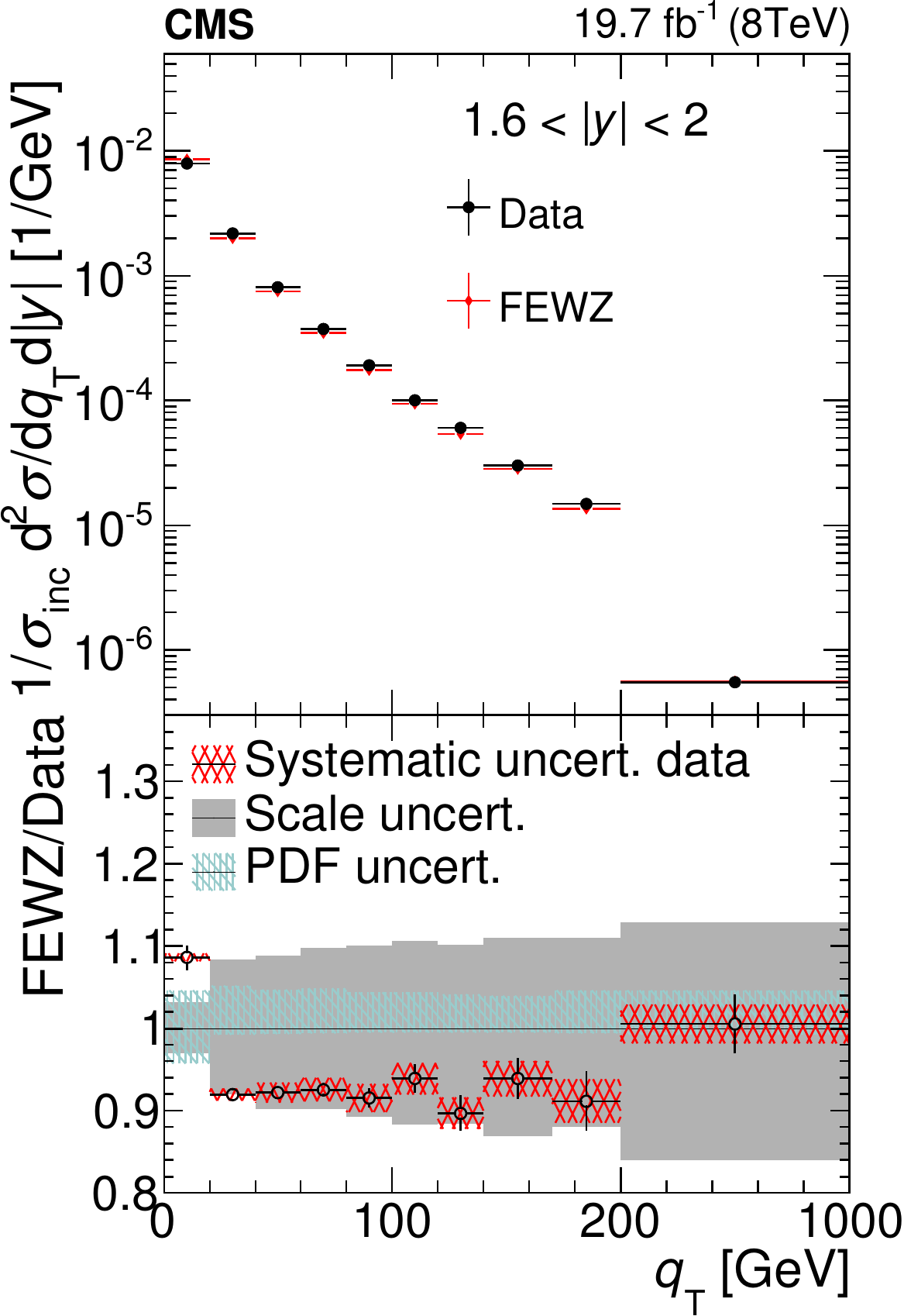}
\includegraphics[width=0.32\textwidth]{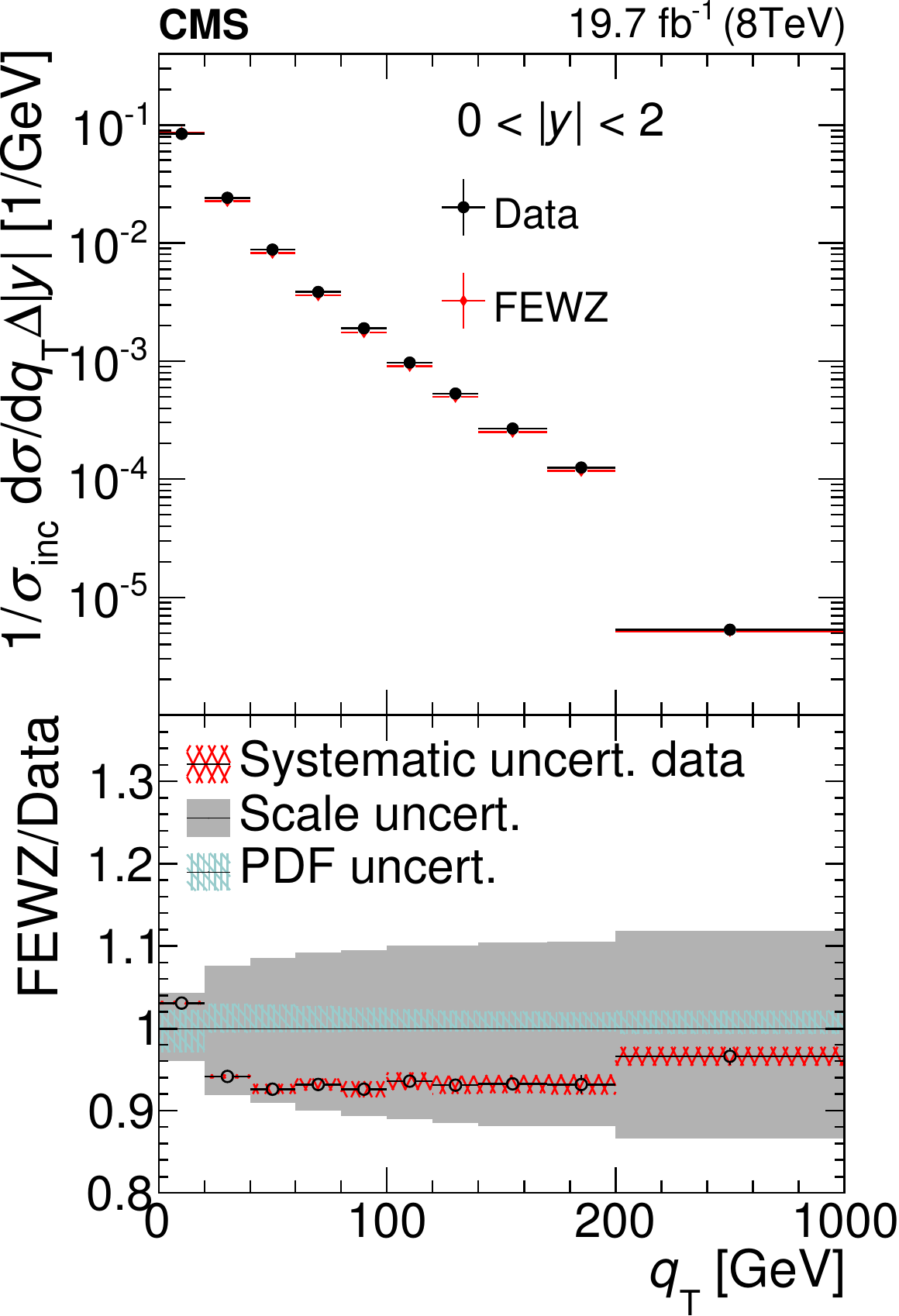}

 \caption{The measured fiducial Z boson differential cross section, normalised to the inclusive fiducial cross section compared to the NNLO prediction of \FEWZ.
The first five plots show the $\qt$ dependence in the five bins of $\abs{y}$ and the last plot shows the
$\qt$ dependence integrated over $\abs{y}$. The NNLO NNPDF23 PDF set with radiative corrections is used for the generation. We include data in $\qt$ up to 1\TeV, but have shortened the bin for presentation purposes.}
\label{fig:dunf_mc1}
\end{figure*}

The double differential cross section normalised to the inclusive
cross section for Z bosons decaying to muons is presented in Table~\ref{tab:XsecY0resultsAndUncNor}.
A comparison of the measurement with the NNLO \FEWZ
computation is shown in Fig.~\ref{fig:dunf_mc1}, where the first five
plots show the $\qt$ dependence in the five bins in $\abs{y}$ and the
last plot shows the $\qt$ dependence integrated over $\abs{y}$. In the
bottom panels the ratio of the \FEWZ prediction to data is shown. The vertical
error bars represent the statistical uncertainties of data and
simulation. The red-hatched bands drawn at the points represent the
systematic uncertainties of the measurement only. The scale
uncertainties are indicated by the grey-shaded areas and the PDF
uncertainties by the light-hatched bands. The scale uncertainties are
estimated from the envelope of the following combinations of variations of
the factorisation $\mu_F$ and the renormalisation $\mu_R$ scales:
($2\mu_F$,$2\mu_R$), ($0.5\mu_F$,$0.5\mu_R$), ($2\mu_F$,$\mu_R$),
($\mu_F$,$2\mu_R$), ($0.5\mu_F$,$\mu_R$), and ($\mu_F$,$0.5\mu_R$). The
PDF uncertainties are evaluated as the envelope of the uncertainties of
the NNLO NNPDF23 and the NNLO CT10~\cite{Gao:2013xoa} PDF sets. The scale uncertainty is
about 4\% for the lowest $\qt$ bin. In the second $\qt$ bin it is
 about 8\% and increases up to about 14\% in the highest $\qt$
bin. The jump in the size of the scale uncertainty between the first
and the second bins in $\qt$ can be understood as a consequence of reducing the order of the calculation
 to NLO when the Z boson is produced in combination with a jet, which is the dominant process for
$\qt>20$\GeV. While the scale uncertainties are smaller at low $\qt$, the shape is not expected to match the data well since multiple soft gluon emissions are not modelled. At very high $\qt$ QED corrections could reach a few percent~\cite{Becher:2013zua,Denner:2011vu}.

 The PDF uncertainties in the region $\qt>20$\GeV range between $+1$\% and
$-4$\%. The uncertainty is asymmetric
 because the inclusive cross section computed using the
NNLO CT10 PDF set is about 2.5\% larger than the one obtained using the NNLO NNPDF23 PDF set.

The NNLO \FEWZ computation predicts the shape correctly, within scale uncertainties of order
6--12\%, where the default scale has the general feature of
underestimating the relative abundance of high-$\qt$ (${>}20\GeV$) events at
the 7\% level. The shape in $\abs{y}$ is well described by \FEWZ.

\begin{figure*}[!htp]
\centering
\includegraphics[width=0.32\textwidth]{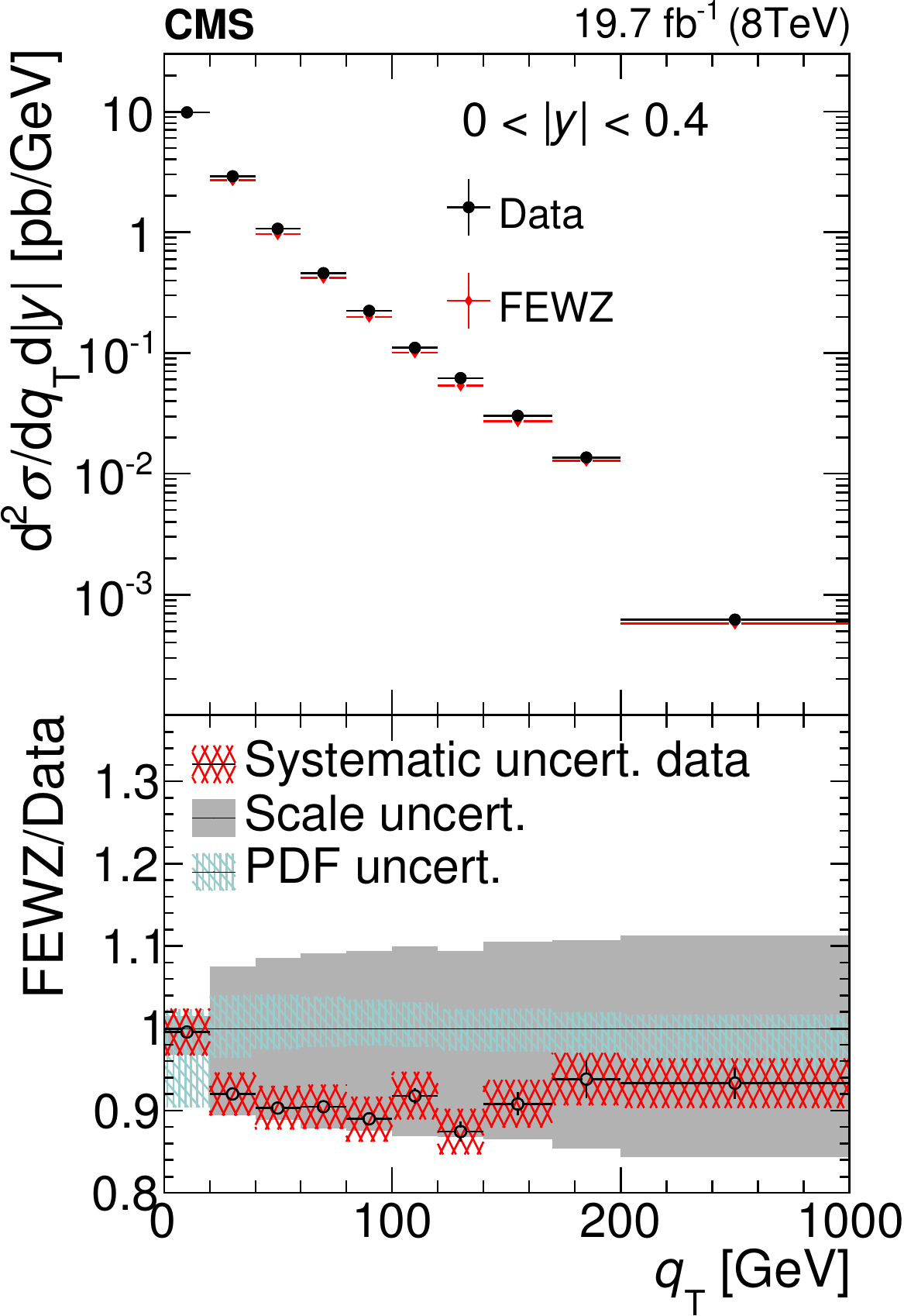}
\includegraphics[width=0.32\textwidth]{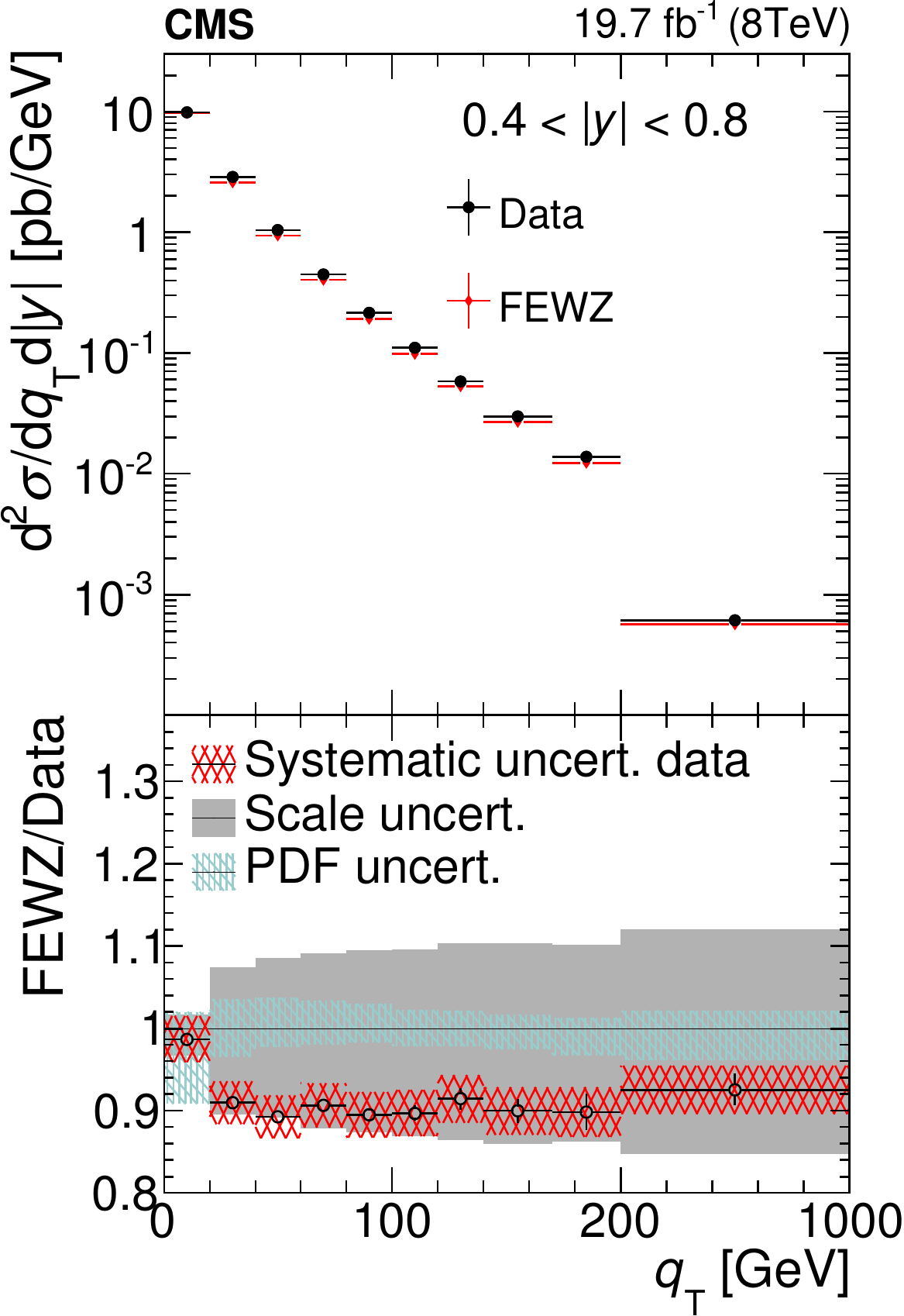}
\includegraphics[width=0.32\textwidth]{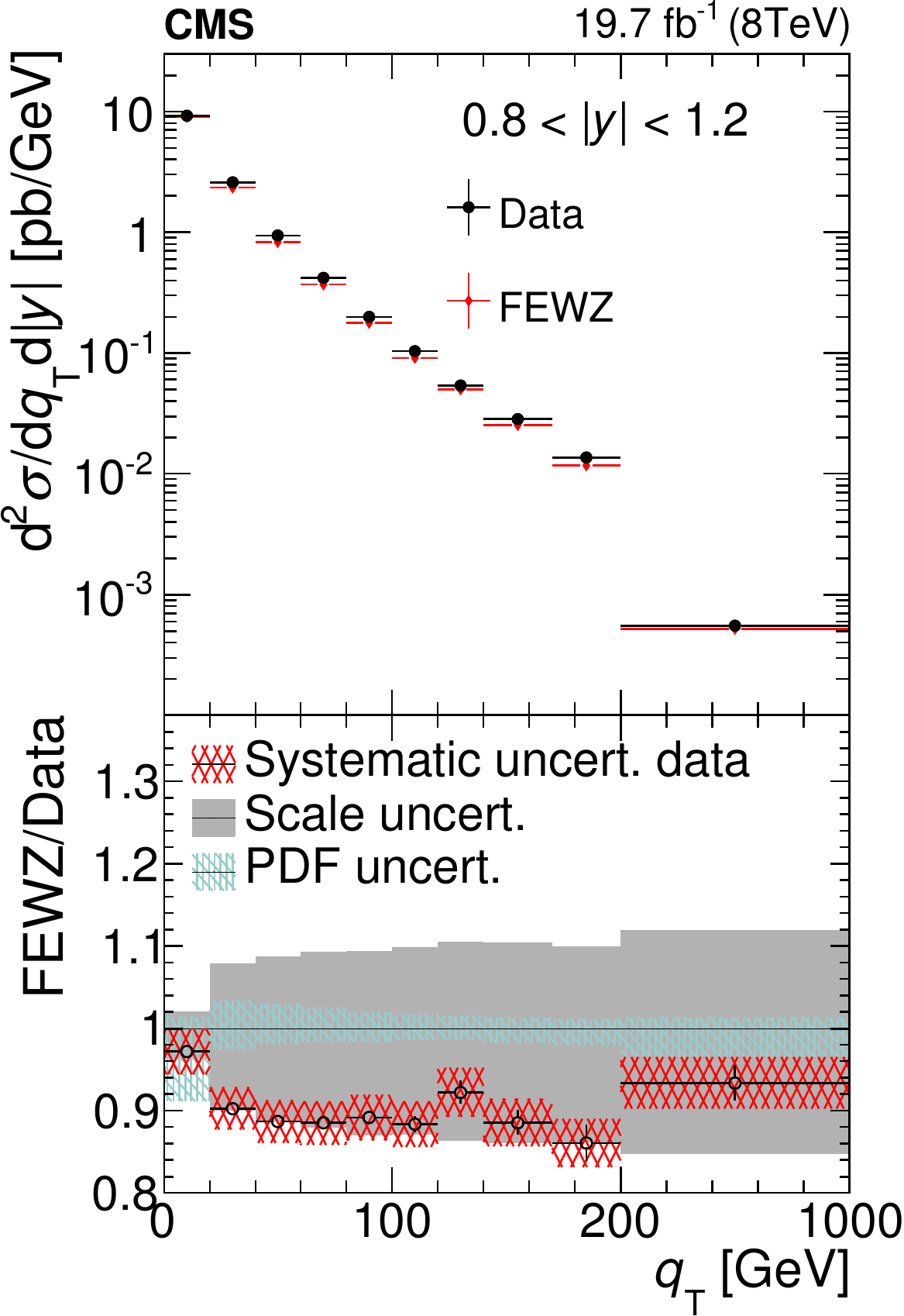}\\
\includegraphics[width=0.32\textwidth]{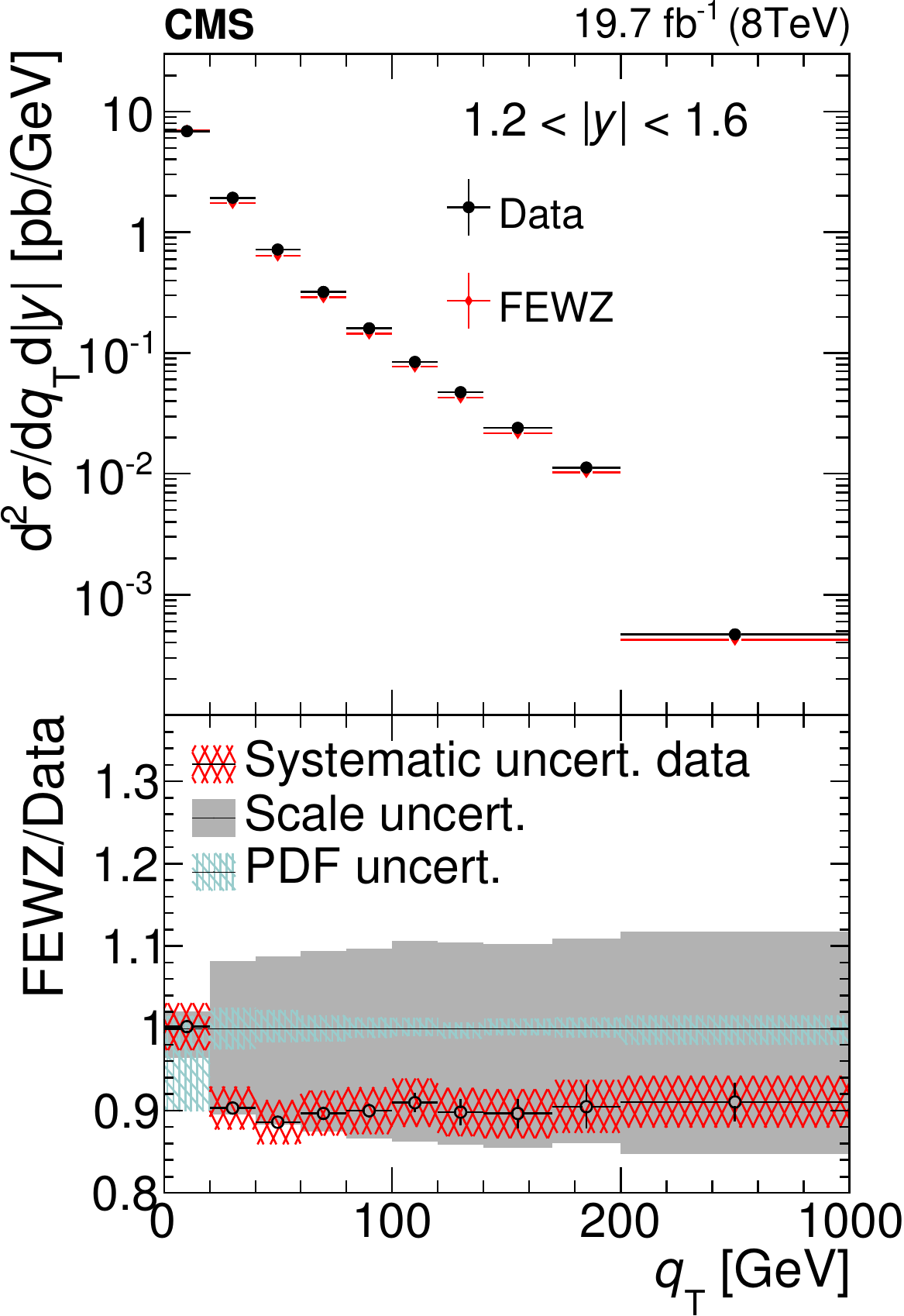}
\includegraphics[width=0.32\textwidth]{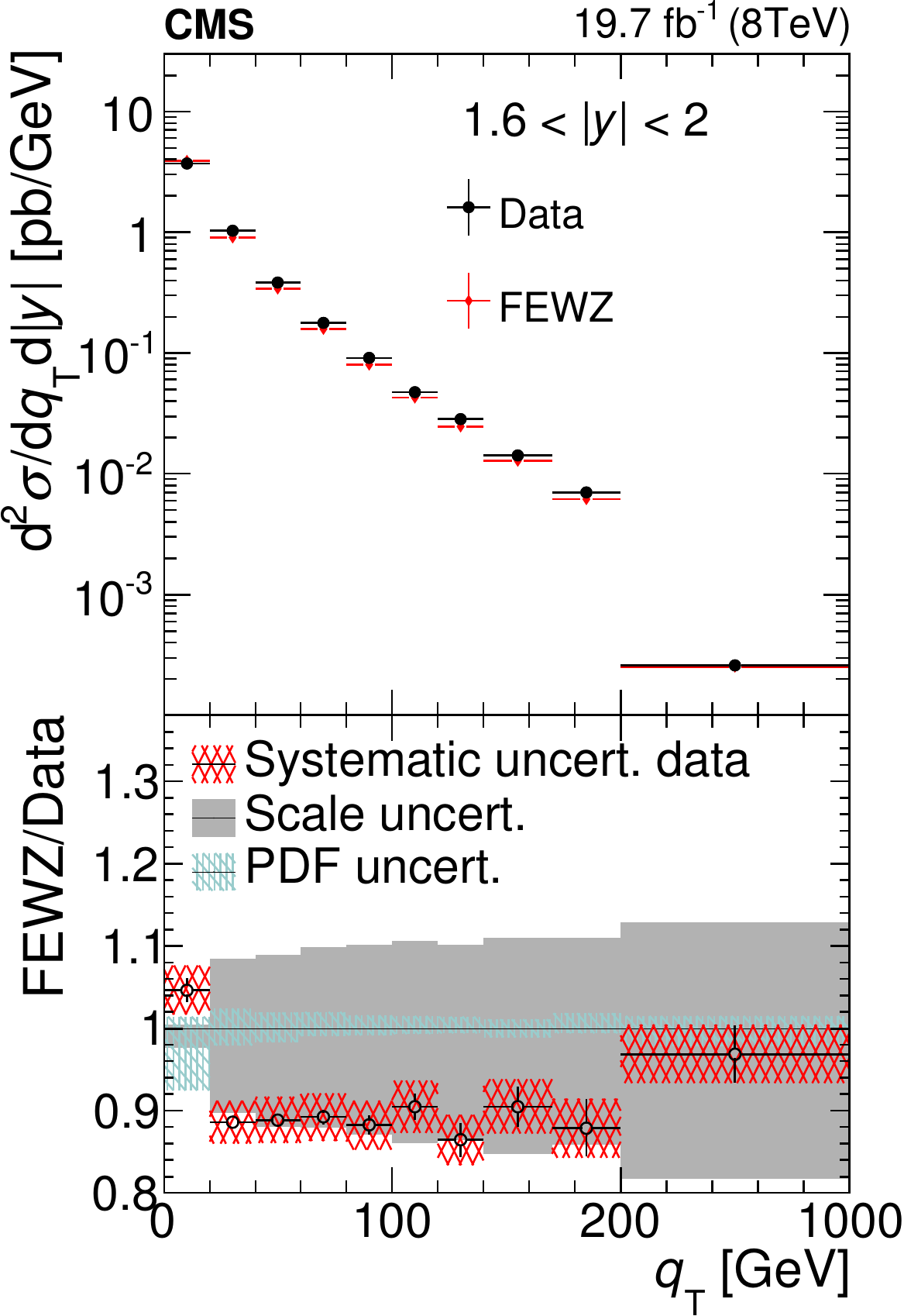}
\includegraphics[width=0.32\textwidth]{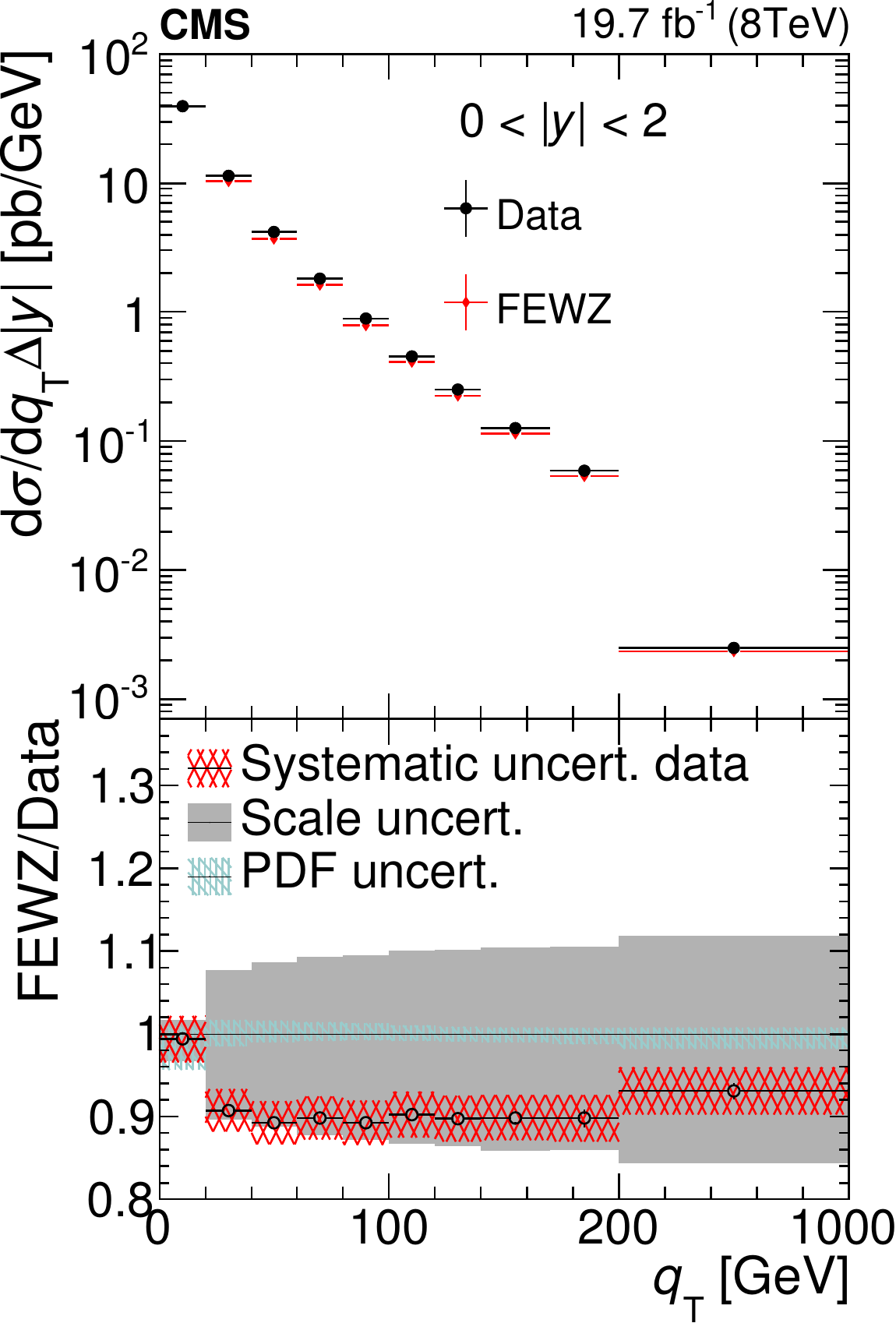}

 \caption{The measured absolute fiducial Z boson differential cross section compared to the NNLO prediction of \FEWZ. The first five plots show the $\qt$ dependence in the five bins of $\abs{y}$ and the last plot shows the $\qt$ dependence integrated over $\abs{y}$. We include data in $\qt$ up to 1\TeV, but have shortened the bin for presentation purposes.}
\label{fig:dunf_abs_mc1}
\end{figure*}

The absolute double differential cross section
 is presented in
Table~\ref{tab:XsecY0resultsAndUncAbs}. The comparison with the NNLO computation of the \FEWZ program is shown in Fig.~\ref{fig:dunf_abs_mc1}. The scale uncertainties range from 10--16\% for $\qt>20$\GeV. The PDF uncertainties are of the order of 3\% in the central rapidity region and decrease to about 1\% in the forward region. The absolute cross section predicted by the NNLO program \FEWZ agrees within the uncertainties with the measurement.

\begin{figure*}[htb]
\centering
\includegraphics[width=0.496\textwidth]{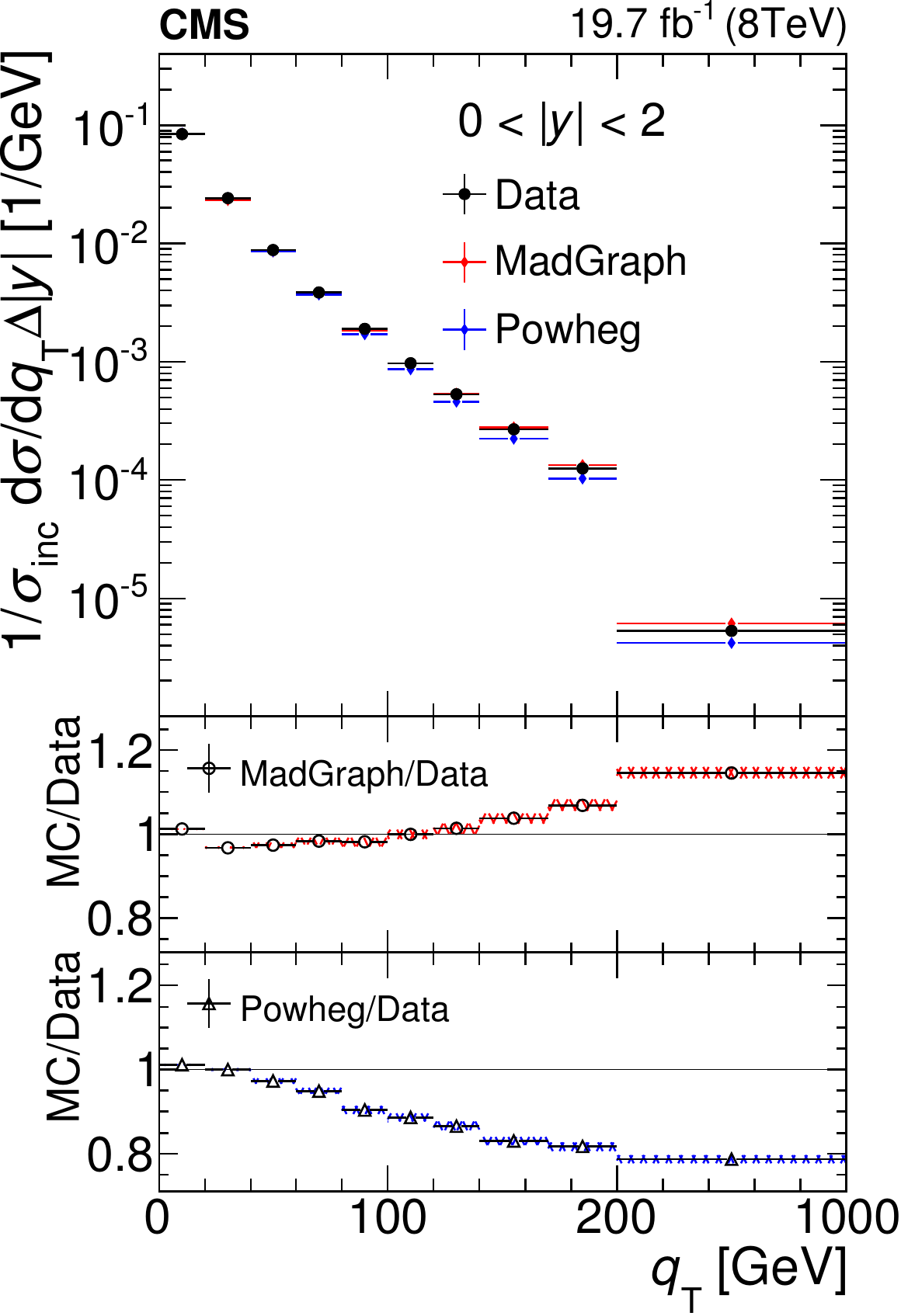}
\includegraphics[width=0.496\textwidth]{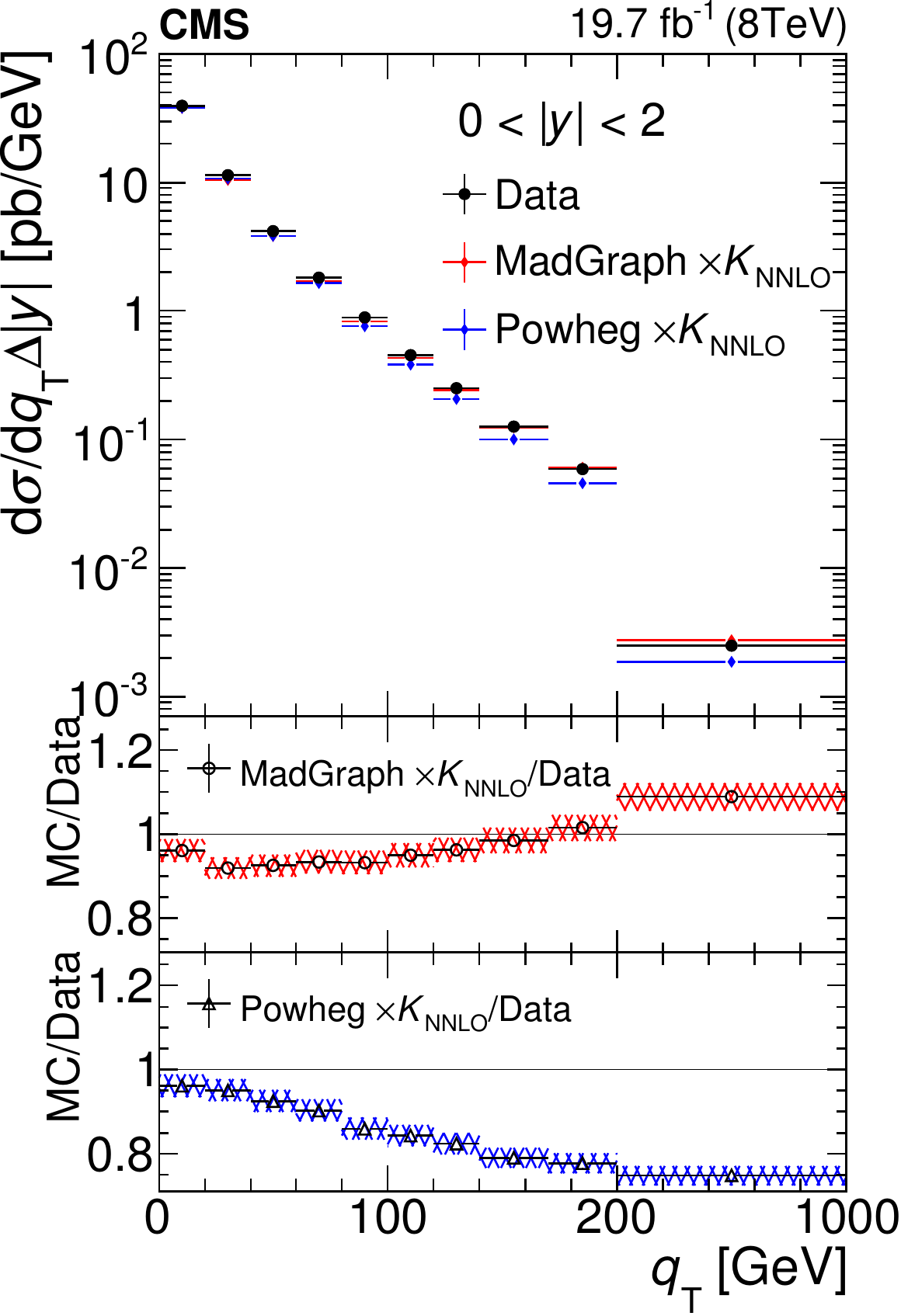}
\caption{Normalised (left) and absolute (right) fiducial Z boson cross section, as a function of $\qt$, compared to predictions from \MADGRAPH (red symbols) and \POWHEG (blue symbols). \MADGRAPH uses the LO CTEQ6L1 PDF set and \POWHEG the NLO CT10 PDF set. The inclusive LO \MADGRAPH and the inclusive NLO \POWHEG cross sections are scaled to the inclusive NNLO cross section calculated with \FEWZ by applying scale factors $K^{FEWZ}_{NNLO}$.}
 \label{fig:dunf_mad_pow}
 \end{figure*}

A comparison of the measurements with the \MADGRAPH and the
\POWHEG generators is shown in Fig.~\ref{fig:dunf_mad_pow}. The statistical uncertainties are smaller than the symbol size. The hatched bands
represent the systematic uncertainties of the measurement only. The two generators show opposite trends in $\qt$. The \MADGRAPH generator overestimates the data in the highest $\qt$ bins, whereas the \POWHEG generator underestimates the data up to 20\% in this region. Also shown are the absolute differential cross section predictions of \MADGRAPH and \POWHEG after normalising their inclusive cross-sections to the NNLO cross section by $K$ factors, that are independent of $\qt$ and $\abs{y}$.

\begin{table*}[ht]
\centering
\topcaption{Measured double differential fiducial cross section normalised to the inclusive fiducial cross section in units of $\GeVns^{-1}$.}\label{tab:XsecY0resultsAndUncNor}
\resizebox{\textwidth}{!}{
\begin{tabular}{c|ccc|ccc|ccc|ccc|ccc} \cline{2-16}
\multicolumn{1}{c}{}& \multicolumn{3}{c|}{$ 0 \leq \abs{y} < 0.4$} & \multicolumn{3}{c|}{$ 0.4 \leq \abs{y} < 0.8$} & \multicolumn{3}{c|}{$ 0.8 \leq \abs{y} < 1.2$} & \multicolumn{3}{c|}{$ 1.2 \leq \abs{y} < 1.6$} & \multicolumn{3}{c}{$ 1.6 \leq \abs{y} < 2$}\\ \hline
\multirow{2}{*}{$\qt$ [\GeVns]} & \multirow{2}{*}{$\rd^2\sigma/\sigma_{\text{inc}}$} & $\delta_{\text{stat}}$ & $\delta_{\text{syst}}$ & \multirow{2}{*}{$\rd^2\sigma/\sigma_{\text{inc}}$} & $\delta_{\text{stat}}$ & $\delta_{\text{syst}}$ & \multirow{2}{*}{$\rd^2\sigma/\sigma_{\text{inc}}$} & $\delta_{\text{stat}}$ & $\delta_{\text{syst}}$ & \multirow{2}{*}{$\rd^2\sigma/\sigma_{\text{inc}}$} & $\delta_{\text{stat}}$ & $\delta_{\text{syst}}$ & \multirow{2}{*}{$\rd^2\sigma/\sigma_{\text{inc}}$} & $\delta_{\text{stat}}$ & $\delta_{\text{syst}}$\\
&&[\%]&[\%]&&[\%]&[\%]&&[\%]&[\%]&&[\%]&[\%]&&[\%]&[\%]\\ \hline
$\mbox{[0,20]}$ & $2.10{\times}10^{-2}$ & 0.09 & 0.30 & $2.10{\times}10^{-2}$ & 0.09 & 0.30 & $1.96{\times}10^{-2}$ & 0.10 & 0.30 & $1.47{\times}10^{-2}$ & 0.12 & 0.31 & $7.88{\times}10^{-3}$ & 0.17 & 0.45  \\
$\mbox{[20,40]}$ & $6.20{\times}10^{-3}$ & 0.18 & 0.44 & $6.08{\times}10^{-3}$ & 0.19 & 0.42 & $5.50{\times}10^{-3}$ & 0.20 & 0.46 & $4.11{\times}10^{-3}$ & 0.24 & 0.59 & $2.17{\times}10^{-3}$ & 0.34 & 0.81  \\
$\mbox{[40,60]}$ & $2.28{\times}10^{-3}$ & 0.30 & 0.84 & $2.22{\times}10^{-3}$ & 0.31 & 0.80 & $1.99{\times}10^{-3}$ & 0.35 & 0.85 & $1.53{\times}10^{-3}$ & 0.40 & 1.03 & $8.11{\times}10^{-4}$ & 0.57 & 1.35  \\
$\mbox{[60,80]}$ & $9.79{\times}10^{-4}$ & 0.47 & 0.99 & $9.48{\times}10^{-4}$ & 0.48 & 0.94 & $8.85{\times}10^{-4}$ & 0.52 & 0.96 & $6.82{\times}10^{-4}$ & 0.62 & 1.16 & $3.75{\times}10^{-4}$ & 0.87 & 1.56  \\
$\mbox{[80,100]}$ & $4.73{\times}10^{-4}$ & 0.69 & 1.33 & $4.56{\times}10^{-4}$ & 0.71 & 1.26 & $4.23{\times}10^{-4}$ & 0.77 & 1.26 & $3.42{\times}10^{-4}$ & 0.89 & 1.43 & $1.92{\times}10^{-4}$ & 1.25 & 1.89  \\
$\mbox{[100,120]}$ & $2.33{\times}10^{-4}$ & 1.02 & 1.44 & $2.34{\times}10^{-4}$ & 1.01 & 1.36 & $2.19{\times}10^{-4}$ & 1.10 & 1.37 & $1.80{\times}10^{-4}$ & 1.25 & 1.50 & $1.01{\times}10^{-4}$ & 1.76 & 2.03  \\
$\mbox{[120,140]}$ & $1.31{\times}10^{-4}$ & 1.37 & 1.51 & $1.24{\times}10^{-4}$ & 1.42 & 1.50 & $1.15{\times}10^{-4}$ & 1.55 & 1.53 & $1.01{\times}10^{-4}$ & 1.72 & 1.58 & $6.03{\times}10^{-5}$ & 2.40 & 2.13  \\
$\mbox{[140,170]}$ & $6.42{\times}10^{-5}$ & 1.57 & 1.59 & $6.34{\times}10^{-5}$ & 1.57 & 1.54 & $6.05{\times}10^{-5}$ & 1.68 & 1.56 & $5.13{\times}10^{-5}$ & 1.93 & 1.68 & $3.00{\times}10^{-5}$ & 2.67 & 2.30  \\
$\mbox{[170,200]}$ & $2.88{\times}10^{-5}$ & 2.36 & 1.91 & $2.93{\times}10^{-5}$ & 2.35 & 1.88 & $2.90{\times}10^{-5}$ & 2.61 & 1.93 & $2.40{\times}10^{-5}$ & 2.98 & 2.14 & $1.49{\times}10^{-5}$ & 4.00 & 2.88  \\
$\mbox{[200,1000]}$ & $1.31{\times}10^{-6}$ & 2.01 & 1.64 & $1.30{\times}10^{-6}$ & 1.96 & 1.57 & $1.17{\times}10^{-6}$ & 2.21 & 1.75 & $9.90{\times}10^{-7}$ & 2.45 & 1.95 & $5.54{\times}10^{-7}$ & 3.39 & 2.39  \\
\hline
\end{tabular}
}
\end{table*}

\begin{table*}[h!tb]
\centering
\topcaption{Measured absolute double differential fiducial cross section in units of pb/\GeVns.}\label{tab:XsecY0resultsAndUncAbs}
\resizebox{\textwidth}{!}{
\begin{tabular}{c|ccc|ccc|ccc|ccc|ccc}
\cline{2-16}
\multicolumn{1}{c}{}& \multicolumn{3}{c|}{$ 0 \leq \abs{y} < 0.4$} & \multicolumn{3}{c|}{$ 0.4 \leq \abs{y} < 0.8$} & \multicolumn{3}{c|}{$ 0.8 \leq \abs{y} < 1.2$} & \multicolumn{3}{c|}{$ 1.2 \leq \abs{y} < 1.6$} & \multicolumn{3}{c}{$ 1.6 \leq \abs{y} < 2$}  \\ \hline
\multirow{2}{*}{$\qt$ [\GeVns]} & \multirow{2}{*}{$\rd^2\sigma$}  & $\delta_{\text{stat}}$ & $\delta_{\text{syst}}$ & \multirow{2}{*}{$\rd^2\sigma$}  & $\delta_{\text{stat}}$ & $\delta_{\text{syst}}$ & \multirow{2}{*}{$\rd^2\sigma$}  & $\delta_{\text{stat}}$ & $\delta_{\text{syst}}$ & \multirow{2}{*}{$\rd^2\sigma$}  & $\delta_{\text{stat}}$ & $\delta_{\text{syst}}$ & \multirow{2}{*}{$\rd^2\sigma$} & $\delta_{\text{stat}}$ & $\delta_{\text{syst}}$ \\
&&[\%]&[\%]&&[\%]&[\%]&&[\%]&[\%]&&[\%]&[\%]&&[\%]&[\%]\\ \hline
$\mbox{[0,20]}$ & 9.87 & 0.10 & 2.84 & 9.86 & 0.10 & 2.85 & 9.20 & 0.10 & 2.85 & 6.89 & 0.12 & 2.85 & 3.71 & 0.18 & 2.87  \\
$\mbox{[20,40]}$ & 2.92 & 0.19 & 2.85 & 2.86 & 0.19 & 2.86 & 2.59 & 0.20 & 2.87 & 1.93 & 0.24 & 2.90 & 1.02 & 0.34 & 2.94  \\
$\mbox{[40,60]}$ & 1.07 & 0.30 & 2.93 & 1.05 & 0.31 & 2.94 & $9.35{\times}10^{-1}$ & 0.35 & 2.97 & $7.19{\times}10^{-1}$ & 0.41 & 3.04 & $3.82{\times}10^{-1}$ & 0.57 & 3.14  \\
$\mbox{[60,80]}$ & $4.61{\times}10^{-1}$ & 0.47 & 2.97 & $4.46{\times}10^{-1}$ & 0.48 & 2.98 & $4.16{\times}10^{-1}$ & 0.52 & 3.00 & $3.21{\times}10^{-1}$ & 0.62 & 3.08 & $1.77{\times}10^{-1}$ & 0.87 & 3.25  \\
$\mbox{[80,100]}$ & $2.23{\times}10^{-1}$ & 0.69 & 3.09 & $2.15{\times}10^{-1}$ & 0.71 & 3.09 & $1.99{\times}10^{-1}$ & 0.77 & 3.12 & $1.61{\times}10^{-1}$ & 0.89 & 3.19 & $9.05{\times}10^{-2}$ & 1.25 & 3.40  \\
$\mbox{[100,120]}$ & $1.10{\times}10^{-1}$ & 1.02 & 3.16 & $1.10{\times}10^{-1}$ & 1.01 & 3.13 & $1.03{\times}10^{-1}$ & 1.10 & 3.15 & $8.46{\times}10^{-2}$ & 1.25 & 3.24 & $4.74{\times}10^{-2}$ & 1.76 & 3.51  \\
$\mbox{[120,140]}$ & $6.18{\times}10^{-2}$ & 1.36 & 3.19 & $5.81{\times}10^{-2}$ & 1.42 & 3.19 & $5.41{\times}10^{-2}$ & 1.55 & 3.22 & $4.76{\times}10^{-2}$ & 1.72 & 3.27 & $2.84{\times}10^{-2}$ & 2.40 & 3.54  \\
$\mbox{[140,170]}$ & $3.02{\times}10^{-2}$ & 1.57 & 3.22 & $2.98{\times}10^{-2}$ & 1.57 & 3.21 & $2.84{\times}10^{-2}$ & 1.69 & 3.24 & $2.41{\times}10^{-2}$ & 1.93 & 3.32 & $1.41{\times}10^{-2}$ & 2.67 & 3.67  \\
$\mbox{[170,200]}$ & $1.36{\times}10^{-2}$ & 2.36 & 3.37 & $1.38{\times}10^{-2}$ & 2.35 & 3.36 & $1.36{\times}10^{-2}$ & 2.61 & 3.43 & $1.13{\times}10^{-2}$ & 2.99 & 3.56 & $7.00{\times}10^{-3}$ & 4.00 & 4.08  \\
$\mbox{[200,1000]}$ & $6.18{\times}10^{-4}$ & 2.01 & 3.24 & $6.12{\times}10^{-4}$ & 1.96 & 3.21 & $5.52{\times}10^{-4}$ & 2.21 & 3.34 & $4.66{\times}10^{-4}$ & 2.45 & 3.44 & $2.60{\times}10^{-4}$ & 3.40 & 3.67  \\ \hline
\end{tabular}
}
\end{table*}

\section{Summary}
For Z bosons decaying to muons the double differential Z boson fiducial cross section in $\qt$ and $\abs{y}$ has been measured in pp collisions at 8\TeV.
The results are compared to the next-to-next-to-leading-order predictions computed with the \FEWZ program and they agree within the scale uncertainties. Deviations from the data of up to 20\% at high transverse momentum are observed in the \MADGRAPH and \POWHEG generators.
The results are presented along with the full covariance matrix in order to enable their use in future fits of the PDF.  The experimental uncertainties are significantly smaller than the current theoretical and PDF uncertainties.

\begin{acknowledgments}
We congratulate our colleagues in the CERN accelerator departments for the excellent performance of the LHC and thank the technical and administrative staffs at CERN and at other CMS institutes for their contributions to the success of the CMS effort. In addition, we gratefully acknowledge the computing centers and personnel of the Worldwide LHC Computing Grid for delivering so effectively the computing infrastructure essential to our analyses. Finally, we acknowledge the enduring support for the construction and operation of the LHC and the CMS detector provided by the following funding agencies: BMWFW and FWF (Austria); FNRS and FWO (Belgium); CNPq, CAPES, FAPERJ, and FAPESP (Brazil); MES (Bulgaria); CERN; CAS, MOST, and NSFC (China); COLCIENCIAS (Colombia); MSES and CSF (Croatia); RPF (Cyprus); MoER, ERC IUT and ERDF (Estonia); Academy of Finland, MEC, and HIP (Finland); CEA and CNRS/IN2P3 (France); BMBF, DFG, and HGF (Germany); GSRT (Greece); OTKA and NIH (Hungary); DAE and DST (India); IPM (Iran); SFI (Ireland); INFN (Italy); MSIP and NRF (Republic of Korea); LAS (Lithuania); MOE and UM (Malaysia); CINVESTAV, CONACYT, SEP, and UASLP-FAI (Mexico); MBIE (New Zealand); PAEC (Pakistan); MSHE and NSC (Poland); FCT (Portugal); JINR (Dubna); MON, RosAtom, RAS and RFBR (Russia); MESTD (Serbia); SEIDI and CPAN (Spain); Swiss Funding Agencies (Switzerland); MST (Taipei); ThEPCenter, IPST, STAR and NSTDA (Thailand); TUBITAK and TAEK (Turkey); NASU and SFFR (Ukraine); STFC (United Kingdom); DOE and NSF (USA).

Individuals have received support from the Marie-Curie program and the European Research Council and EPLANET (European Union); the Leventis Foundation; the A. P. Sloan Foundation; the Alexander von Humboldt Foundation; the Belgian Federal Science Policy Office; the Fonds pour la Formation \`a la Recherche dans l'Industrie et dans l'Agriculture (FRIA-Belgium); the Agentschap voor Innovatie door Wetenschap en Technologie (IWT-Belgium); the Ministry of Education, Youth and Sports (MEYS) of the Czech Republic; the Council of Science and Industrial Research, India; the HOMING PLUS program of the Foundation for Polish Science, cofinanced from European Union, Regional Development Fund; the Compagnia di San Paolo (Torino); the Consorzio per la Fisica (Trieste); MIUR project 20108T4XTM (Italy); the Thalis and Aristeia programs cofinanced by EU-ESF and the Greek NSRF; and the National Priorities Research Program by Qatar National Research Fund.
\end{acknowledgments}

\bibliography{auto_generated}
\cleardoublepage \appendix\section{The CMS Collaboration \label{app:collab}}\begin{sloppypar}\hyphenpenalty=5000\widowpenalty=500\clubpenalty=5000\textbf{Yerevan Physics Institute,  Yerevan,  Armenia}\\*[0pt]
V.~Khachatryan, A.M.~Sirunyan, A.~Tumasyan
\vskip\cmsinstskip
\textbf{Institut f\"{u}r Hochenergiephysik der OeAW,  Wien,  Austria}\\*[0pt]
W.~Adam, T.~Bergauer, M.~Dragicevic, J.~Er\"{o}, M.~Friedl, R.~Fr\"{u}hwirth\cmsAuthorMark{1}, V.M.~Ghete, C.~Hartl, N.~H\"{o}rmann, J.~Hrubec, M.~Jeitler\cmsAuthorMark{1}, W.~Kiesenhofer, V.~Kn\"{u}nz, M.~Krammer\cmsAuthorMark{1}, I.~Kr\"{a}tschmer, D.~Liko, I.~Mikulec, D.~Rabady\cmsAuthorMark{2}, B.~Rahbaran, H.~Rohringer, R.~Sch\"{o}fbeck, J.~Strauss, W.~Treberer-Treberspurg, W.~Waltenberger, C.-E.~Wulz\cmsAuthorMark{1}
% ...
\vskip\cmsinstskip
% ...
1:~~Also at Vienna University of Technology, Vienna, Austria\\
2:~~Also at CERN, European Organization for Nuclear Research, Geneva, Switzerland\\
% ...

\end{sloppypar}
\end{document}